\documentclass{article}
\usepackage[top = 4cm, bottom = 4cm, left = 3cm, right = 3cm]{geometry}
\usepackage[utf8]{inputenc}
\usepackage{adjustbox}
\usepackage{algorithm}
\usepackage{algorithmicx}
\usepackage{algpseudocode}
\usepackage{amsmath}
\usepackage{amssymb}
\usepackage{amsthm}

\usepackage{bm}
\usepackage{braket}
\usepackage{chemfig}
\usepackage{color}
\usepackage{dsfont}
\usepackage{bm}
\usepackage{makecell}
\usepackage{parskip}
\usepackage{subcaption}
\usepackage{graphicx}
\usepackage{url}

\usepackage[maxbibnames=99]{biblatex} 
\definecolor{darkgreen}{rgb}{0,0.4,0}

\newcommand{\am}[1]{{\color{blue} #1}}

\graphicspath{ {./figures/} }

\begin{document}

\title{Peptide conformational sampling using the Quantum Approximate Optimization Algorithm}
\author{Sami Boulebnane\thanks{University College London and Phasecraft Ltd; {\tt sami.boulebnane.19@ucl.ac.uk}.},
\
Xavier Lucas\thanks{Roche Pharma Research and Early Development, Roche Innovation Center Basel, F. Hoffmann-La Roche Ltd., CH-4070 Basel, Switzerland},
\
Agnes Meyder\thanks{Roche Pharma Research and Early Development, Roche Innovation Center Basel, F. Hoffmann-La Roche Ltd., CH-4070 Basel, Switzerland}, \\
\
Stanislaw Adaszewski\thanks{Roche Pharma Research and Early Development, Roche Innovation Center Basel, F. Hoffmann-La Roche Ltd., CH-4070 Basel, Switzerland; {\tt stanislaw.adaszewski@roche.com}.},
\
and Ashley Montanaro\thanks{Phasecraft Ltd and University of Bristol; {\tt ashley@phasecraft.io}.}}
\maketitle


\begin{abstract}
    Protein folding --- the problem of predicting the spatial structure of a protein given its sequence of amino-acids --- has attracted considerable research effort in biochemistry in recent decades.
    In this work, we explore the potential of quantum computing to solve a simplified version of protein folding. More precisely, we numerically investigate the performance of a variational quantum algorithm, the Quantum Approximate Optimization Algorithm (QAOA), in sampling low-energy conformations of short peptides. We start by benchmarking the algorithm on an even simpler problem: sampling self-avoiding walks, which is a necessary condition for a valid protein conformation. Motivated by promising results achieved by QAOA on this problem, we then apply the algorithm to a more complete version of protein folding, including a simplified physical potential.
    In this case, based on numerical simulations on 20 qubits, we find less promising results: deep quantum circuits are required to achieve accurate results, and the performance of QAOA can be matched by random sampling up to a small overhead.
    Overall, these results cast serious doubt on the ability of QAOA to address the protein folding problem in the near term, even in an extremely simplified setting. We believe that the approach and conclusions presented in this work could offer valuable methodological insights on how to systematically evaluate variational quantum optimization algorithms on real-world problems beyond protein folding.
\end{abstract}

\section{Introduction}

Protein folding has been a major focus in biochemistry and computational research in recent decades, motivated by its central role in protein function and protein homeostasis \cite{clausen_abildgaard_gersing_stein_lindorff-larsen_hartmann-petersen_2019,nassar_dignon_razban_dill_2021}. From a computational perspective, protein structure prediction methods can be broadly classified into knowledge-based and physics-based approaches, along with more recent deep learning algorithms \cite{pearce_zhang_2021}.  Indeed, the field has experienced a revolution with the publication of AlphaFold2, an AI algorithm capable of predicting the 3D apo structure of single protein domains and multimeric systems with experimental accuracy~\cite{jumper_evans_pritzel_green_figurnov_ronneberger_tunyasuvunakool_bates_zidek_potapenko_et_al_2021,evans_o_neill_pritzel_antropova_senior_green_zidek_bates_blackwell_yim_et_al_2021}. Nonetheless, further advances in the field could benefit de novo protein structure design, improve accurate protein structure prediction for low homology proteins, and provide better understanding of intrinsically disordered protein regions \cite{woolfson_2021,ruff_pappu_2021,shea_best_mittal_2021}. In this regard, the potential of quantum computing algorithms to vastly and unbiasedly explore protein conformational ensembles is currently under investigation by us and others~\cite{robert_barkoutsos_woerner_tavernelli_2021}.

There exists a significant body of research on applying quantum algorithms to protein dynamics~\cite{Babbush2014,PerdomoOrtiz2012,1810.13411,1811.00713,Robert2021,2105.09690,2110.08163,2110.01589}. While \cite{2105.09690} considers the resource requirements to run a Monte Carlo simulation of a protein fragment on a quantum computer and \cite{2110.08163,2110.01589} introduce hybrid classical-quantum algorithms to estimate the energy of protein complexes, the vast majority of these works address the \textit{combinatorial} formulation of protein folding. Given a (classical) potential function expressing the energy of the protein from the positions of all its atoms, the latter problem consists of predicting the minimum energy conformation \cite{alberts2002molecular}. This optimization problem in many continuous variables may be simplified to a discrete problem: lattice protein folding, whereby atom or amino acid positions are restricted to a lattice \cite{lattice_model_protein}. Other approximations include using simplified potential functions; we refer to e.g. \cite{lattice_protein_folding_2018} for a comprehensive review. Lattice-based methods were convenient approaches to protein folding in the early age of classical computing, when resources were too scarce to produce meaningful results with the original problem formulation. Due to even more restricted resources on real quantum computers and even on quantum emulators, lattice models are also the primary focus in the quantum computing literature \cite{alberts2002molecular}. Lattice-based protein folding is a computationally hard optimization problem \cite{CRESCENZI1998,BERGER1998}, making it a target of interest for quantum optimization algorithms.

\subsection{Our contribution}
In this work, similarly to \cite{1810.13411,1811.00713}, we consider applying the Quantum Approximate Optimization Algorithm (QAOA) to a lattice-based protein folding problem. Following \cite{Robert2021}, the protein is discretized on a tetrahedral lattice to approximate a realistic geometry with a limited number of qubits. The choice of QAOA is motivated by its expected scalability based on its good trainability properties \cite{2008.02941,2109.06957}. This distinguishes this study from \cite{Robert2021} which uses a problem-independent ansatz; such variational circuits are known to pose several challenges, including barren plateaus \cite{cost_function_dependent_barren_plateaus_2021} and the existence of many spurious local minima \cite{2109.06957}, making them less well-adapted to large-scale problems.  Our work also differs from all earlier contributions in that the protein is modeled at the atomic level (and not as a sequence of amino acids) and attributed a physically realistic potential (see section \ref{sec:hamiltonian_evolution}), as done in classical conformer generators. Such an approach is only viable for a fault-tolerant quantum computer as it requires costly arithmetic circuits and generally long-range qubit interactions. Therefore, while earlier studies focused on resource estimation and optimization with near-term applicability in view and eschewed the question of the performance and scalability of the algorithm towards practically relevant problem instances, we take the opposite stance: assuming access to a sufficiently powerful quantum computer to compute the problem's cost function to arbitrary precision and optimize a variational circuit with an arbitrary number of layers, we consider the performance and limitations of QAOA in this idealized setting. This allows us to better understand the potential performance of QAOA for instances beyond the capability of classical methods. Finally, unlike earlier contributions that focused on finding minimum-energy peptide conformations, we propose to examine the distribution of solutions sampled from QAOA beyond their expected energy. Such an approach may be desirable in future searches for a quantum advantage via QAOA, since its superiority over classical optimization algorithms has until now remained elusive, but it provably\footnote{Under plausible complexity theory conjectures.} achieves quantum supremacy when used as a sampler \cite{1602.07674}.

\subsection{Summary of results}
\begin{figure}[!tbp]
    \centering
    \includegraphics[width=0.8\textwidth]{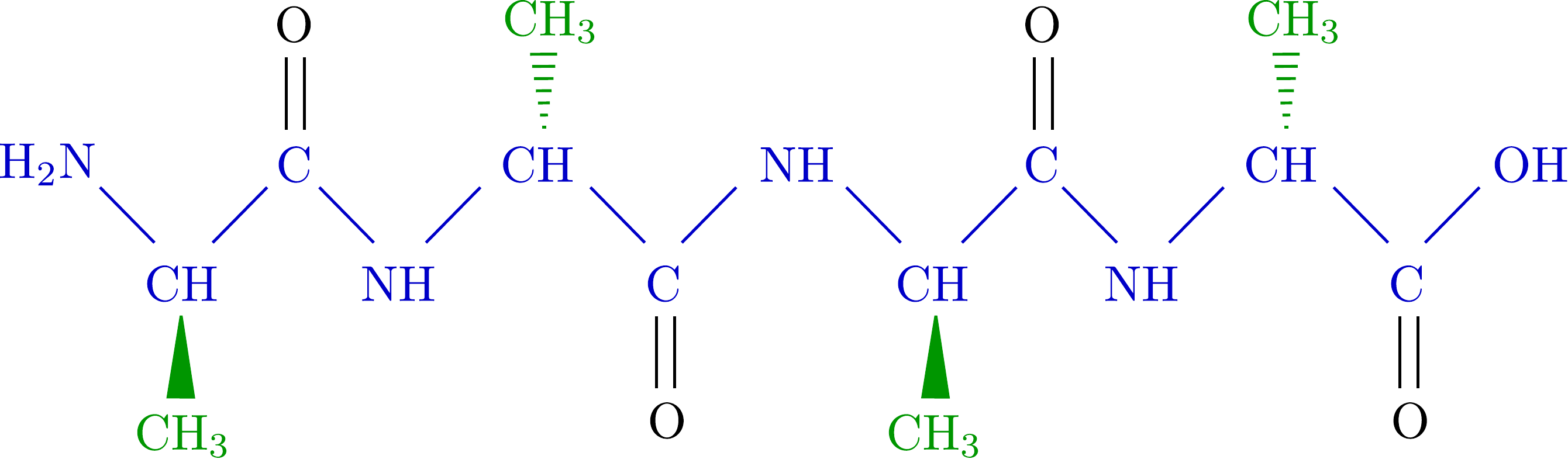}
    \caption{Alanine tetrapeptide (blue: backbone chain; green: side chains)}
    \label{fig:ala_tetrapeptide}
\end{figure}

First, we consider applying QAOA to a relaxed version of the protein folding problem. A protein discretized on a lattice can be thought of as an instance of a self-avoiding walk on the lattice, so in order to produce a physically realistic configuration, QAOA must output a self-avoiding walk. Therefore, one can simplify the protein folding problem and render it more mathematically tractable by considering the problem of just outputting a self-avoiding walk on a lattice using QAOA. Although this problem is trivially solvable using a classical algorithm, it can provide insight into the expected performance of QAOA for the more realistic protein folding problem. We numerically show that QAOA can be efficiently variationally optimized to produce self-avoiding walks with non-trivial probability; more precisely, for fixed ansatz depth, the probability of sampling a self-avoiding walk appears to improve over uniformly random sampling by a factor increasing exponentially with the problem size. Besides, for the limited-scale experiments reported here (up to 28 qubits), an ansatz depth $p \sim 10$ consistently produces a self-avoiding walk with probability at least $10\,\%$, meaning a $10^4$ improvement for the largest walk size considered. Unfortunately, owing to the limited size of instances accessible to classical simulation, it remains unclear how the ansatz depth should scale with the problem size to achieve a constant success probability.

Next, we apply QAOA to a small lattice protein folding problem in the Lennard-Jones model. Due to the limited number of qubits, we are not able to encode the positions of all atoms of the peptide. To address this difficulty, we choose to only represent heavy atoms from the backbone chain, integrating out all other atoms. The cost function to minimize, depending only on the heavy atoms from the backbone chain, is then taken to be the Lennard-Jones potential, partially minimized over configurations of the side chains and light atoms of the backbone chain. We stress that this partial minimization procedure is a mere artifact designed to make numerical experiments feasible; it is a priori not scalable and would not be required to implement the algorithm on a large enough real quantum computer. We use numerical simulations on 20 qubits to address an alanine peptide (see figure \ref{fig:ala_tetrapeptide}), which is a common benchmark for classical molecular dynamics simulation algorithms, see e.g. \cite{wang_zhang_wu_lei_cieplak_duan_2006}. We compare several different initialization methods. We determine that QAOA seems to find this more realistic (though still highly simplified) protein folding significantly more challenging. In order to achieve an expected energy close to the minimum (within relative error $\sim10^{-2}$) more than 40 ansatz layers are required. In addition, we analyse the ability of random guessing to simulate QAOA, in the sense of sampling from those configurations which QAOA obtains with high probability. We find that sampling using a QAOA ansatz with $p\in\{2,3,8,62\}$ layers can be matched by fewer than $6p$ random guesses.

Despite the mixed nature of these results, we believe that the present work outlines methods that could in principle be applied to benchmark QAOA on any other optimization problem. Besides, while earlier research laid a lot of emphasis on the circuit implementation of QAOA when addressing protein folding or other real-world problems, our findings suggest that the major difficulty with the algorithm may not relate to the quantum circuit but rather to the expressivity and/or practical trainability of the ansatz. From a methodological standpoint, this hints that the last two issues should be clarified first when looking for problems where QAOA offers advantage.

\subsection{Related work}
Various earlier works attempted to apply quantum optimization algorithms or heuristics to lattice formulations of the protein folding problem. Babbush et al.~\cite{Babbush2014} introduced several formulations of protein folding as an optimization problem on boolean variables, making it suitable for both quantum and classical SAT solvers. Besides, the authors described two encodings of the problem: the turn-based encoding (used in the rest of this paper) and the diamond encoding, the latter being more qubit-greedy. More precisely, while the turn-based encoding describes the chains of the peptide as a sequence of turns (each turn being encoded on a constant number of qubits), the diamond-based encoding locates each atom on a discretized sphere. As chains grow longer, so do the spheres on which atoms may lie, hence the higher number of qubits required per atom. \cite{Babbush2014} also proposed two techniques to penalize overlaps of the folded protein: the first uses quantum arithmetic to compute auxiliary boolean variables recording clashes, while the second resorts to slack variables representing distances between all pairs of atoms. Perdomo-Ortiz et al. \cite{PerdomoOrtiz2012} applied the ideas of the latter work to fold small proteins (up to 6 amino-acids) on a square lattice using a 80-qubit quantum annealer. More recently, Fingerhuth et al. \cite{1810.13411}, based on the same earlier work, considered using a unary turn-based encoding on a cubic lattice. The latter encoding simplified the expression of the problem's Hamiltonian, at the cost of requiring more qubits and a non-trivial QAOA mixer (see section \ref{sec:qaoa}). The authors managed to run the algorithm on a real quantum computer and obtain non-trivial results for a 3 amino acids protein (tripeptide). Later on, Babej et al. \cite{1811.00713} improved the complexity of the arithmetic in \cite{1810.13411} and folded a 10 residues protein on a quantum annealer.

Very recently, Robert et al. \cite{Robert2021} proposed to use a tetrahedral lattice (unlike square or cubic lattices from previous works) to ensure more realistic interatomic bond angles despite the coarseness of the discretized geometry\footnote{However, it must be underlined that a tetrahedral lattice does not cover all types of bond angles found in nature. For instance, the restriction to a tetrahedral lattice does not allow to faithfully represent planar bonds, e.g. \chemfig{C=O}. In the present work, these bonds are still properly accounted for as only part of the protein, excluding planar bonds in particular, is represented on the lattice.}. However, due to limited resources on real quantum computers, the authors ultimately did not attempt a modelization at the atomic level and, in line with previous works, described the protein as a sequence of amino acids instead. An original proposal of the work is to implement pair interactions using a complex system of penalty terms, removing the need for quantum arithmetic. Unfortunately, the method requires a number of qubits scaling exponentially in the range of the interactions accounted in the model, practically limiting the experiments to nearest-lattice-neighbour interactions. Finally, the work distinguished itself from previous studies by using a problem-independent, hardware-optimized ansatz to look for a solution. This removed the need for coherently implementing the problem's Hamiltonian on the quantum computer. However, the scalability of the approach remains open given the well-known empirical \cite{1805.12037} and theoretical \cite{cost_function_dependent_barren_plateaus_2021,2109.06957} challenges plaguing problem-independent ansätze. Here, our focus is instead on the use of a problem-dependent ansatz which, although it may perform less well on a small-scale system, is expected to have superior scalability properties.

\section{The Quantum Approximate Optimization Algorithm}
\label{sec:background}
\label{sec:qaoa}

We start by giving a brief overview of the main quantum optimization algorithm used in this work --- the Quantum Approximate Optimization Algorithm~\cite{1411.4028} (QAOA). QAOA is a variational quantum algorithm~\cite{2012.09265} designed to (approximately) minimize a classical cost function $H(x)$ depending on $n$ binary variables $x \in \{0, 1\}^n$.

In its simplest version (see e.g.~\cite{1709.03489} for more sophisticated approaches designed to handle constraints among other improvements), QAOA starts with a quantum state $\ket{\Psi_0}$ representing a uniform superposition of all solutions: 
\begin{align}
    \ket{\Psi_0} & := \frac{1}{\sqrt{2^n}}\sum_{x \in \{0, 1\}^n}\ket{x},
\end{align}
where the $j^{\textrm{th}}$ bit $x_j$ of $x$ represents the value of the $j^{\textrm{th}}$ binary variable. QAOA alternates application to $\ket{\Psi_0}$ of a Hamiltonian evolution under the \textit{problem Hamiltonian}:
\begin{align}
    H_C & := \sum_{x \in \{0, 1\}^n}H(x)\ket{x}\bra{x}
\end{align}
and a \textit{mixer Hamiltonian}
\begin{align}
    H_B & := \sum_{j \in [n]}X_j,
\end{align}
where $X_j$ is the Pauli $X$ matrix acting on the $j^{\textrm{th}}$ qubit. The evolution times are hyperparameters to be optimized classically. The quantum state prepared by \textit{level-$p$ QAOA} explicitly reads:
\begin{align}
    \ket{\Psi_{\textrm{QAOA}}(\bm\beta, \bm\gamma)} & := U_B(\beta_{p - 1})U_C(\gamma_{p - 1})\ldots U_B(\beta_0)U_C(\gamma_0)\ket{+}^{\otimes n},
\end{align}
where
\begin{align}
    U_C(\theta) & := \exp\left(-\frac{i\theta H_C}{2}\right),\\
    U_B(\theta) & := \exp\left(-\frac{i\theta H_B}{2}\right)
\end{align}
are the time evolution operators associated to $H_C$, $H_B$ and the parameters $\bm\beta, \bm\gamma \in \mathbf{R}^p$, commonly referred to as \textit{QAOA angles}, are the corresponding evolution times. Once $\ket{\Psi_{\textrm{QAOA}}(\bm\beta, \bm\gamma)}$ has been prepared, it can be measured in the computational basis, yielding a bitstring $x \in \{0, 1\}^n$ with cost $H(x)$; by Born's rule, the expected cost of this bitstring is $\braket{\Psi_{\textrm{QAOA}}|H_C|\Psi_{\textrm{QAOA}}}$ and the angles are classically optimized to minimize this expectation. Once optimal (or good enough) $\bm\beta, \bm\gamma$ have been found and provided $\braket{\Psi_{\textrm{QAOA}}|H_C|\Psi_{\textrm{QAOA}}}$ is sufficiently close to $\min\limits_{x \in \{0, 1\}^n}H(x)$, good solutions to the original optimization problem can be obtained by measuring $\ket{\Psi_{\textrm{QAOA}}(\bm\beta, \bm\gamma)}$ in the computational basis. In this work, the variational parameters are always optimized using the L-BFGS algorithm from NLopt~\cite{Johnson2011}.

\section{Sampling self-avoiding walks with QAOA}
\label{sec:qaoa_saw}
A folded peptide, modeled as a ramified chain of atoms or amino acids, can be regarded as a particular instance of a self-avoiding walk. Therefore, a classical or quantum conformer sampler should only generate configurations satisfying this constraint. Unfortunately, it is already non-trivial whether a quantum algorithm can achieve that with sufficiently high probability. Therefore, in this section, we include no interaction at all and simply investigate the capability of QAOA to sample self-avoiding walks. More precisely, we consider training a QAOA ansatz to sample self-avoiding walks without any requirement on their distribution, contrary to lattice-based protein folding where the attractive interaction potential is to be minimized among valid (self-avoiding) configurations. Therefore, the performance of QAOA on the former task may plausibly provide an upper bound on its capabilities when applied to lattice-based protein folding.

\subsection{Description of the problem}
A self-avoiding walk on a lattice (see e.g. \cite{1206.2092} for a detailed introduction) is a path on the lattice where each site is visited at most once. Self-avoiding walks are well-studied in statistical physics and find applications in polymer physics among others fields; however, rigorous mathematical results remain scarce. The theory of self-avoiding walks has occasionally been explicitly applied to protein folding~\cite{Bahi2013}.

\begin{figure}[!tbp]
\centering
\begin{subfigure}{0.3\textwidth}
    \centering
    \includegraphics[width=0.8\textwidth]{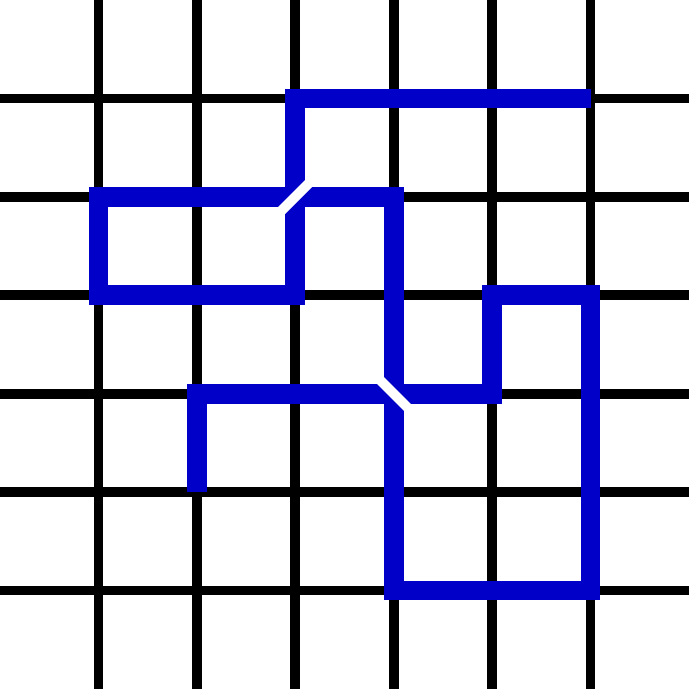}
    \caption{Self-crossing walk}
\end{subfigure}
\begin{subfigure}{0.3\textwidth}
    \centering
    \includegraphics[width=0.8\textwidth]{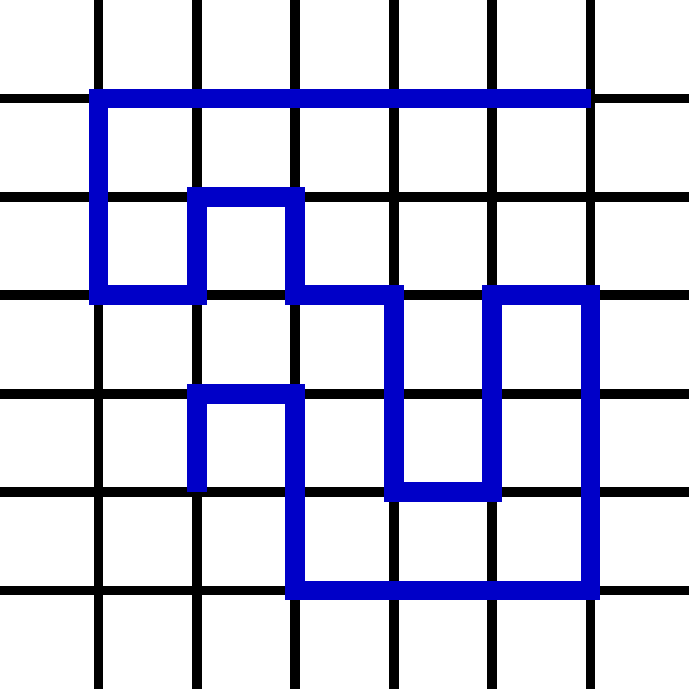}
    \caption{Self-avoiding walk}
\end{subfigure}
\begin{subfigure}{0.3\textwidth}
    \centering
    \includegraphics[width=0.8\textwidth]{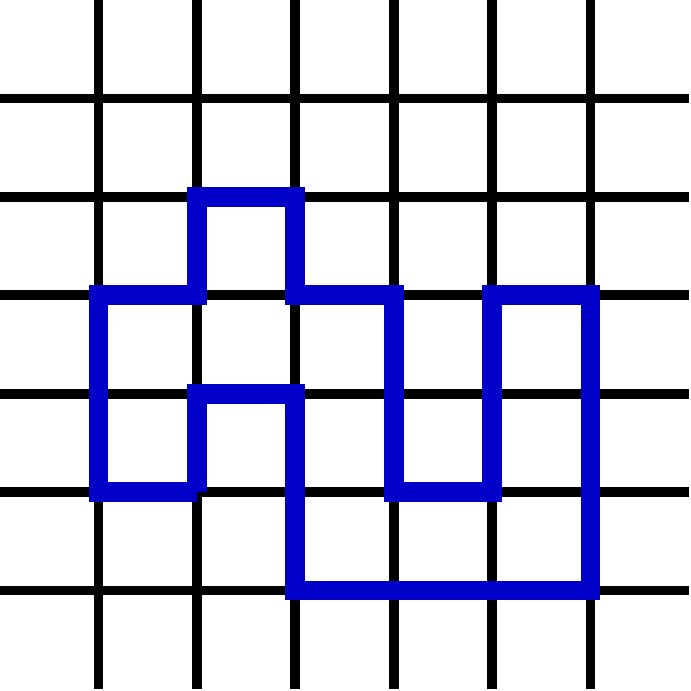}
    \caption{Self-avoiding loop}
\end{subfigure}
\caption{Walks on a square lattice}
\label{fig:saw_examples}
\end{figure}

In this work, we only consider the self-avoiding walk on $\mathbf{Z}^2$ (two-dimensional square lattice). This choice is motivated by the limited number of lattice directions to encode (4, corresponding to 2 qubits) and the fact that self-avoiding walks are, from numerical simulations, rarer on a square lattice than on the tetrahedral lattice used in section \ref{sec:sampling_protein_conformations_qaoa}. The scarcity of valid walks ensures that the problem is sufficiently hard and allows to better understand the scaling of the success probability with the number of layers in the quantum variational ansatz. In order to further decrease the number of valid walks for a given walk length, we further enforce the condition that the path be a loop (the start and end points coincide). This constraint can be physically motivated when modelling an antibody loop for instance. Illustrations are given in figure \ref{fig:saw_examples}.

\subsection{QAOA implementation}
\label{sec:saw_qaoa_implementation}
We now describe the implementation of a self-avoiding walk sampler using QAOA.

\subsubsection{Encoding and Hamiltonian}
\label{sec:qaoa_saw_encoding_hamiltonian}
A walk on the square lattice is represented by a turn-based encoding as described in section \ref{sec:encoding_generalities}. We considered both the absolute (4 directions) and the relative (3 directions) encodings.

Concerning the choice of classical Hamiltonian to be minimized via QAOA, one requires a function which is easily implementable as a quantum circuit and minimal iff.\ the configuration is a self-avoiding loop. We propose to use the Hamiltonian (in a schematic form)
\begin{align}
\label{eq:sal_hamiltonian}
    H & = \textrm{number of self crossings} + \lambda \times \left(\textrm{distance between walk endpoints}\right)^2,
\end{align}
where $\lambda > 0$ is a penalty coefficient. It is easily seen that the second term can be implemented using $\mathcal{O}\left(n^2\right)$ two-qubit $Z$ rotations $\hat{R}_{zz}(\theta) := e^{\frac{i\theta}{2}Z_iZ_j}$ and depth $\mathcal{O}(n)$. More precisely, we can encode each turn in two qubits, where the first qubit represents the horizontal direction and the second qubit the vertical one. The variation of horizontal and vertical coordinates between the endpoints of the walk can then each be expressed as a sum of $Z_i$. Therefore, the distance squared between the endpoints will be a sum of terms $Z_iZ_j$. The first term is more challenging and can for instance be implemented by computing the pairwise distances between all sites of the walk. This can be done with $\mathcal{O}\left(n^2\right)$ gates and depth $\mathcal{O}\left(n\log n\right)$ using efficient quantum arithmetic \cite{1805.12445}. In both cases, the depth grows slower than the number of sites $\Omega\left(n^2\right)$ as operations can be parallelized between all pairs $(i, j)$; this is achieved by graph edge colouring, whereby a colouring with $n$ colours can be efficiently computed classically \cite{Misra1992}.

Alternatively, one may have chosen a boolean function taking value $0$ iff.\ the walk is a self-avoiding walk and $1$ otherwise. However, arguments similar to the optimality proof of Grover's unstructured search algorithm \cite{Ambainis2002} show that in this case, this basic quantum algorithm is guaranteed to perform at least as well as QAOA. In fact, QAOA will even have a strictly higher runtime as it requires a classical optimization of the ansatz parameters, which Grover does not. In contrast, the optimality proof of Grover's algorithm does not carry over to Hamiltonians with more than two energies and therefore leaves open the possibility of sampling good configurations with high probability by applying QAOA to Hamiltonian (\ref{eq:sal_hamiltonian}), including in constant depth.

\subsubsection{Mixer}
\label{sec:saw_mixers}
While restricting to the Hamiltonian defined in section \ref{sec:qaoa_saw_encoding_hamiltonian}, we considered several candidates for the QAOA mixer (see section \ref{sec:qaoa}): the standard QAOA mixer on $n$ qubits $e^{-\frac{i\beta}{2}\sum_{k \in [n]}X_k}$, the qudit QAOA mixer introduced in \cite{2011.13420} and described in section \ref{sec:sampling_protein_conformations_qaoa} $\sum_{j \in [d]}e^{-\frac{i\beta^{(j)}}{2}}Z_d^{j}\ket{+}\bra{+}\left(Z_d^j\right)^{\dagger}$ and an ``inversion about the mean" mixer $e^{-\frac{i\beta}{2}\ket{+}\bra{+}}$. Since the qudit mixer is a generalization of the ``inversion about the mean" one, the former should always outperform the latter; however, it also requires to optimize more parameters per layer.

Note that the standard qubit QAOA mixer is only applicable when the number of encoded turns is a power of two, excluding the relative turn-based encoding (3 encoded turns). The mixer $e^{-\frac{i\beta}{2}\ket{+}\bra{+}} = 1 + \left(e^{-\frac{i\beta}{2}} - 1\right)\ket{+}\bra{+}$, which is similar to an inversion about the mean operator $\mathbf{1}_{2^n} - 2\ket{+}_n\bra{+}_n$, is exactly equivalent\footnote{Precisely, \cite{1810.13411} uses a unary encoding of the turns, while this work uses a binary encoding. The SWAP mixer from \cite{1810.13411} acts on the unary encoding the same way as the inversion about the mean mixer does on the binary encoding.} to the mixer based on SWAP operators proposed in \cite{1810.13411}.

\subsection{Experiments}
\label{sec:saw_experiments}
The variational parameters of the QAOA ansatz described in section \ref{sec:saw_qaoa_implementation} were optimized for the two choices (relative vs. absolute) of configuration encoding and three choices of mixers (standard qubit mixer, qudit mixer, ``inversion about the mean" mixer). We considered walks up to 10 steps (encoded on 16 qubits).

The penalty coefficient $\lambda$ in the problem Hamiltonian (\ref{eq:sal_hamiltonian}) was tuned so as to maximize the probability of sampling a valid configuration for variational parameters minimizing the energy. More precisely, we optimized QAOA at levels $p \in \{1, 2, 3\}$ for different values of $\lambda$ and selected the $\lambda$ maximizing the probability of a valid configuration for optimal QAOA angles; we then used this fixed penalty coefficient for optimization with further layers. The procedure is illustrated in figure \ref{fig:penalty_tuning_6_steps} (resp. \ref{fig:penalty_tuning_10_steps}) from appendix \ref{appendix:figures_saw_qaoa} for a 6-step (resp. 10-step) walk; it suggests an optimal value of $0.2-0.3$ for the penalty coefficient. To facilitate the parameter search, the $\bm\gamma$ QAOA angles were rescaled so that the expected energy achieved by QAOA varied according to the same length scale in $\bm\beta$ and $\bm\gamma$: the adequate rescaling was determined from the level $p = 1$ optimization landscape and assumed to carry over to higher depth.

After fixing the penalty coefficient to $\lambda = 0.2$, the QAOA ansatz for the resulting Hamiltonian was optimized at levels $1 \leq p \leq 10$, starting in each case from $2000$ angles $\bm\beta, \bm\gamma$ angles drawn uniformly from $[-2\pi, 2\pi]^p, [-10\pi, 10\pi]^p$ respectively (it is sufficient to restrict to these intervals given the energies are multiple of $\lambda = 0.2$). Concurrently to this simple random initialization method for the ansatz, we considered initializing variational parameters at depth $p$ from extrapolating optimal parameters found at previous levels, as proposed in \cite{Zhou2020}. In this work, we simply used a linear extrapolation and applied the method from $p = 5$. Precisely, this means that for optimal level-$p$ parameters $\bm{\beta^{*(p)}}, \bm{\gamma^{*(p)}}$, level-$(p + 1)$ parameters $\bm{\beta^{(p + 1)}}, \bm{\gamma^{(p + 1)}}$ are initialized as:
\begin{align}
    \left\{\begin{array}{cccc}
        \beta^{(p + 1)}_j & \longleftarrow & \beta^{*(p)}_j & 0 \leq j < p\\
        \gamma^{(p + 1)}_j & \longleftarrow & \gamma^{*(p)}_j & 0 \leq j < p\\
        \beta^{(p + 1)}_p & \longleftarrow & 2\beta^{*(p)}_{p - 1} - \beta^{*(p)}_{p - 2} \\
        \gamma^{(p + 1)}_p & \longleftarrow & 2\gamma^{*(p)}_{p - 1} - \gamma^{*(p)}_{p - 2} 
    \end{array}\right..
\end{align}

\subsection{Results}
\label{sec:saw_results}
As an example, we report some results obtained for the 10-step self-avoiding walk in tables \ref{tab:mixer_comparison_backtracking} and \ref{tab:mixer_comparison_non_backtracking}; more data points are included on figure \ref{fig:mixer_comparison} from appendix \ref{appendix:figures_saw_qaoa}. The probability $P_{\textrm{SAW}}$ of sampling a self-avoiding walk from the QAOA state is given against the ansatz depth for the 3 possible choices of mixers and 2 possible choices of encodings. Besides, QAOA is compared with the simpler quantum amplitude amplification algorithm \cite{Brassard2002}. More precisely, since depth $p$ QAOA makes $p$ queries to the optimization problem's cost function, it can naturally be compared to amplitude amplification with $p$ queries to the oracle.
\begin{table}[!tbp]
	\centering
	\begin{tabular}{c|c|c|c|c}
		Algorithm | depth & 0 & 2 & 5 & 10\\
		\hline
		QAOA - inversion about mean & 0.000671 & 0.0206 & 0.104 & 0.403\\
		\hline
		QAOA - qudit $X$ mixer & 0.000671 & 0.0207 & 0.0721 & *\\
		\hline
		QAOA - qubit $X$ mixer & 0.000671 & 0.0160 & 0.0694 & 0.202\\
		\hline
		Amplitude amplification & 0.000671 & 0.0167 & 0.0791 & 0.268
	\end{tabular}
	\caption{Success probability for backtracking encoding}
	\label{tab:mixer_comparison_backtracking}
\end{table}

\begin{table}[!tbp]
	\centering
	\begin{tabular}{c|c|c|c|c}
		Algorithm | depth & 0 & 2 & 5 & 10\\
		\hline
		QAOA - inversion about mean & 0.00671 & 0.0527 & 0.129 & 0.357\\
		\hline
		QAOA - qudit $X$ mixer & 0.00671 & 0.0529 & 0.130 & *\\
		\hline
		Amplitude amplification & 0.00671 & 0.158 & 0.615 & 0.977
	\end{tabular}
	\caption{Success probability for non-backtracking encoding}
	\label{tab:mixer_comparison_non_backtracking}
\end{table}

\begin{figure}[!tbp]
	\centering
    \begin{subfigure}{0.43\textwidth}
        \includegraphics[width=\textwidth]{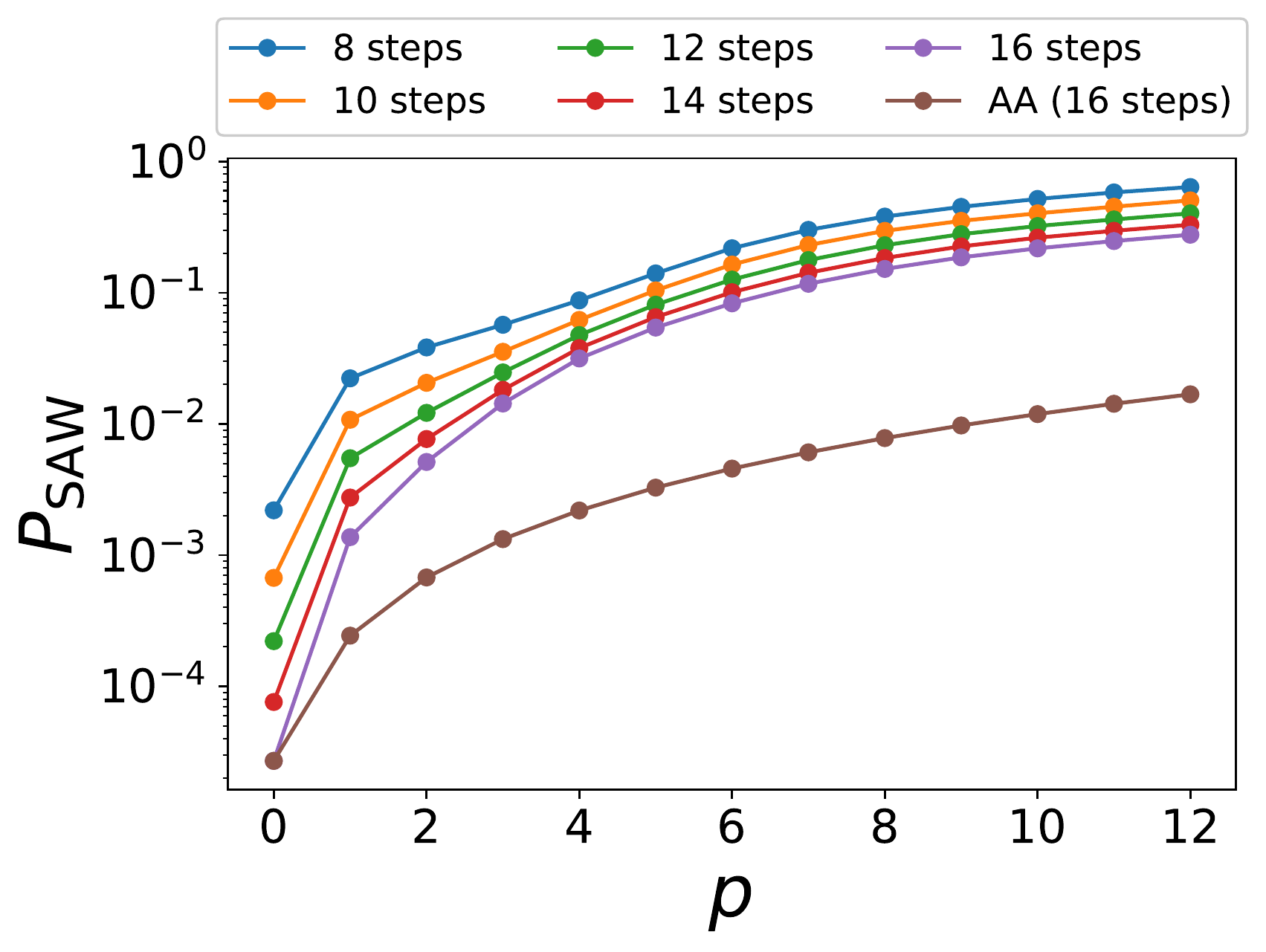}
    	\caption{Probability of self-avoiding loop for different number of steps. For reference, the amplitude amplification result is also represented for the 16-steps walk.}
    	\label{fig:success_probability_vs_p}
    	\vspace*{50px}
    \end{subfigure}
    \hspace*{0.05\textwidth}
    \begin{subfigure}{0.47\textwidth}
        \includegraphics[width=\textwidth]{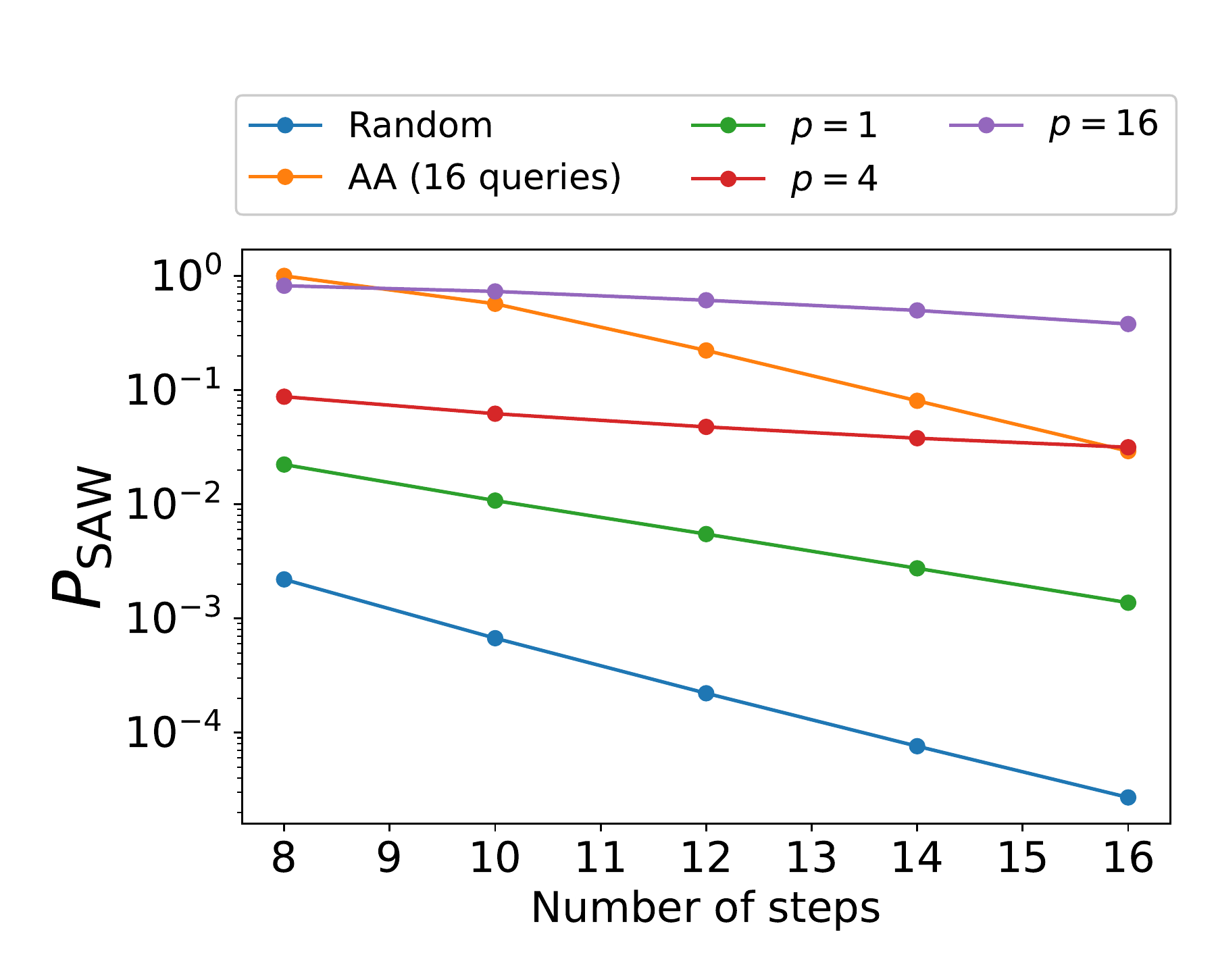}
    	\caption{Probability of self-avoiding loop vs. problem size, for different $p$. For reference, amplitude amplification with 16 queries (comparable to $p = 16$ QAOA) is also represented. Asymptotically, the amplitude curve should be parallel to the random one, which can already be observed for the small problem sizes represented. Exponential fits: $y = 10^{-0.776 - 0.238x}$ (random), $y = 10^{-0.453 - 0.151x}$ ($p = 1$), $y = 10^{-0.642 - 0.0550x}$ ($p = 4$), $y = 10^{0.270 - 0.0418x}$ ($p = 16$). Correlation coefficients are all $> 99\,\%$.}
    	\label{fig:success_probability_vs_problem_size}
    \end{subfigure}
    \caption{Probability of QAOA sampling a self-avoiding loop as a function of ansatz depth and problem size.}
\end{figure}

One first observes that the non-backtracking encoding achieves a better performance than the backtracking one at $p = 2$, since the success probability compared to random assignment ($p = 0$) is multipled by $8$ in the former case and only $3$ in the latter. This justifies our choice for this encoding for the alanine tetrapeptide problem in section \ref{sec:sampling_protein_conformations_qaoa}. Besides, the results suggest that the qudit mixer only marginally outperforms the ``inversion about the mean" one despite being harder to optimize. The latter is illustrated by the $p = 5$ results from table \ref{tab:mixer_comparison_backtracking}, where the qudit mixer underperforms the ``inversion about the mean", while the contrary holds for optimal variational parameters (see section \ref{sec:saw_mixers}). This implies that the best variational parameters found for the qudit mixer after $2000$ optimization attempts are still suboptimal. Therefore, the gain in success probability given by the qudit mixer does not seem to justify the extra optimization cost. For the 10-step walk example detailed here, QAOA does not visibly outperform amplitude amplification. If this proved correct, this would pose a serious challenge to QAOA as amplitude amplification merely provides a quadratic speed-up over random guessing \cite{Brassard2002} while self-avoiding walks are exponentially rare among all lattice walks (see e.g. \cite{DuminilCopin2012} for an explicit example). We therefore compare the performance of both algorithms for larger problem instances --- up to 16 steps, corresponding to 28 qubits --- fixing the encoding (absolute turn-based encoding) and the mixer (inversion about the mean). Figure \ref{fig:success_probability_vs_p} shows the success probability against the ansatz depth $p$. In all cases, the probability appears to increase exponentially for small $p$ (contrasting with the quadratic increase of amplitude amplification) before saturating. In appendix \ref{appendix:figures_saw_qaoa}, we provide a more detailed analysis of this success probability for a 16-step walk.
To confirm that QAOA behaves qualitatively differently from amplitude amplification, we also explicitly considered the success probability at low fixed depth for increasing problem size. The results are reported in figure \ref{fig:success_probability_vs_problem_size} and suggest QAOA improves over random assignment with a ratio growing exponentially with the problem size, while this improvement would be constant for amplitude amplification.

\begin{figure}[!tbp]
    \centering
    \begin{subfigure}{0.4\textwidth}
        \centering
        \includegraphics[width=\textwidth]{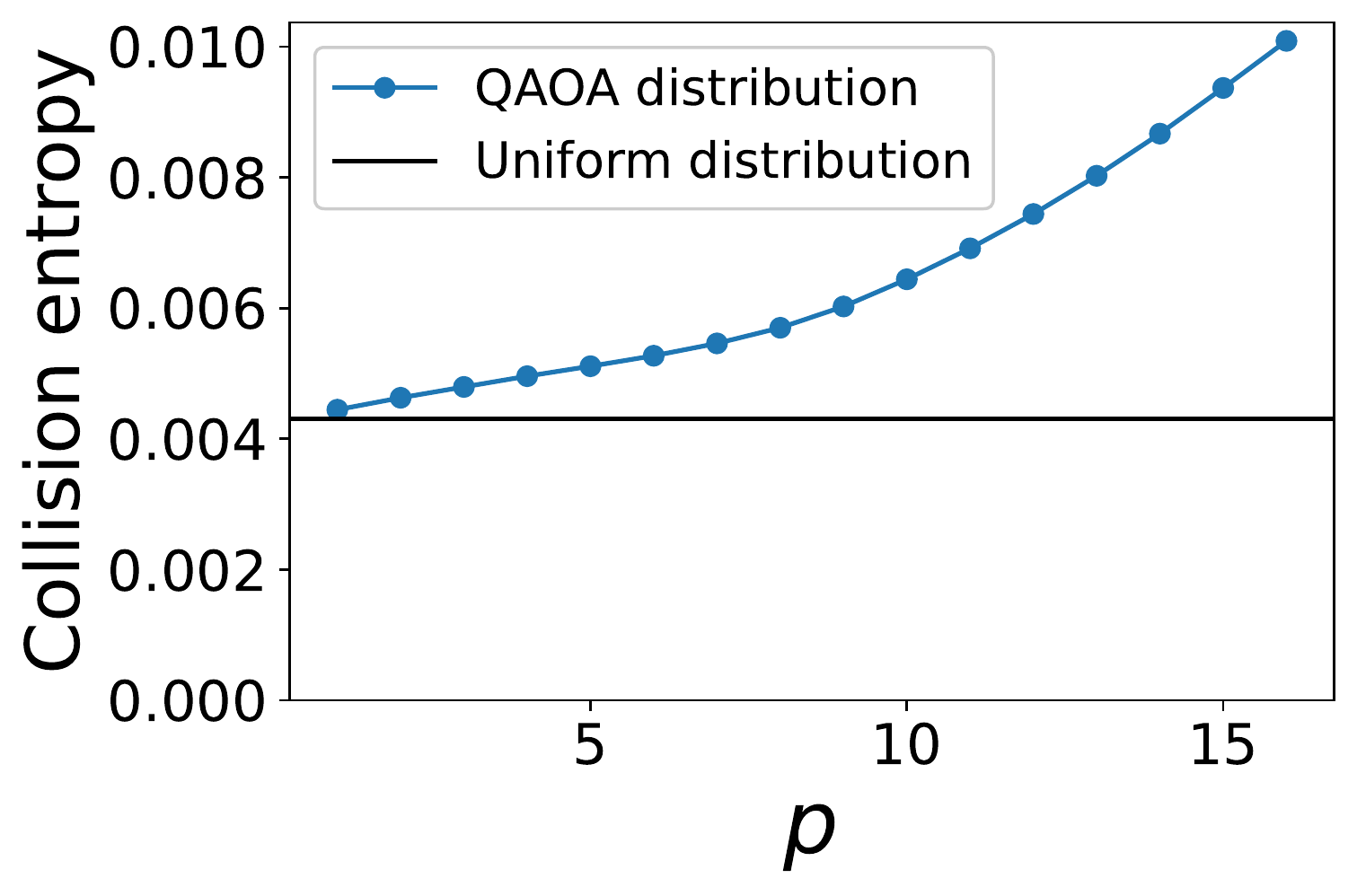}
        \caption{Collision entropy}
    \end{subfigure}
    \hspace*{0.1\textwidth}
    \begin{subfigure}{0.4\textwidth}
        \centering
        \includegraphics[width=\textwidth]{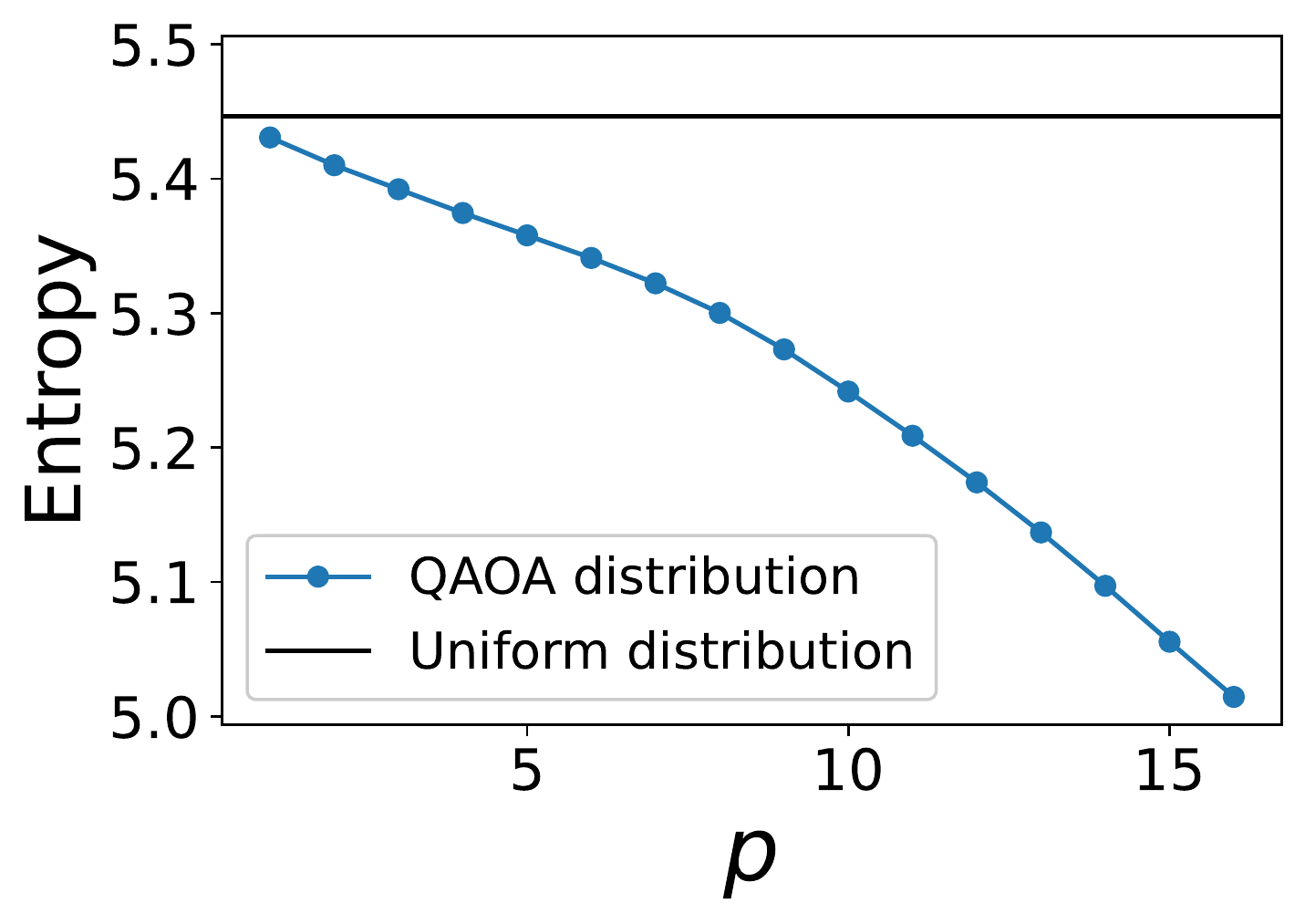}
        \caption{Shannon entropy}
    \end{subfigure}
    \caption{Entropy of self-avoiding walk distribution sampled from QAOA for increasing number of layers (12-steps walk, inversion-about-the-mean mixer, absolute turn-based encoding)}
    \label{fig:saw_distribution_entropies}
\end{figure}

Another important question is whether QAOA, when it succeeds, samples \textit{fairly} from the set of self-avoiding walks. Classically, Monte Carlo algorithms (see e.g. \cite{monte_carlo_methods_saw_2009} for a detailed review) have been known for decades to sample self-avoiding walks. Although these algorithms are usually heuristic, it seems possible to derive theoretical guarantees of (almost) fair sampling at least for restricted cases~\cite{randall_2000}. In this study, we quantify the uniformity of the sampling by considering the collision entropy and the standard Shannon entropy. Given the probability distribution $\left(p_x\right)_{x \in \mathcal{X}}$ of a random variable $X$ taking values in a set $\mathcal{X}$, the collision entropy of $X$ is defined as:
\begin{align}
    \textrm{CollisionEntropy}(X) & := \sum_{x \in \mathcal{X}}p_x^2.
\end{align}
It is $\frac{1}{|\mathcal{X}|}$ iff.\ $X$ is uniformly distributed and $1$ iff.\ $X$ is deterministic. The Shannon entropy is defined as:
\begin{align}
    H(X) & := -\sum_{x \in \mathcal{X}}p_x\log p_x
\end{align}
and is $\log|X|$ iff. $X$ is uniformly distributed, $0$ iff. $X$ is deterministic. Both quantities are represented in figure \ref{fig:saw_distribution_entropies} for the 12-steps self-avoiding walk. They remain close to their expected values for a uniform distribution, although the situation degrades with increasing $p$. Since by figure \ref{fig:success_probability_vs_p} $p$ is required to increase with the number of steps to achieve a constant success probability, it is difficult to infer how this property will persist at larger problem sizes. However, one can say qualitatively that QAOA is sampling from a random, but nonuniform, distribution.

\begin{figure}[!tbp]
    \centering
    \begin{subfigure}{0.32\textwidth}
        \centering
        \includegraphics[width=\textwidth]{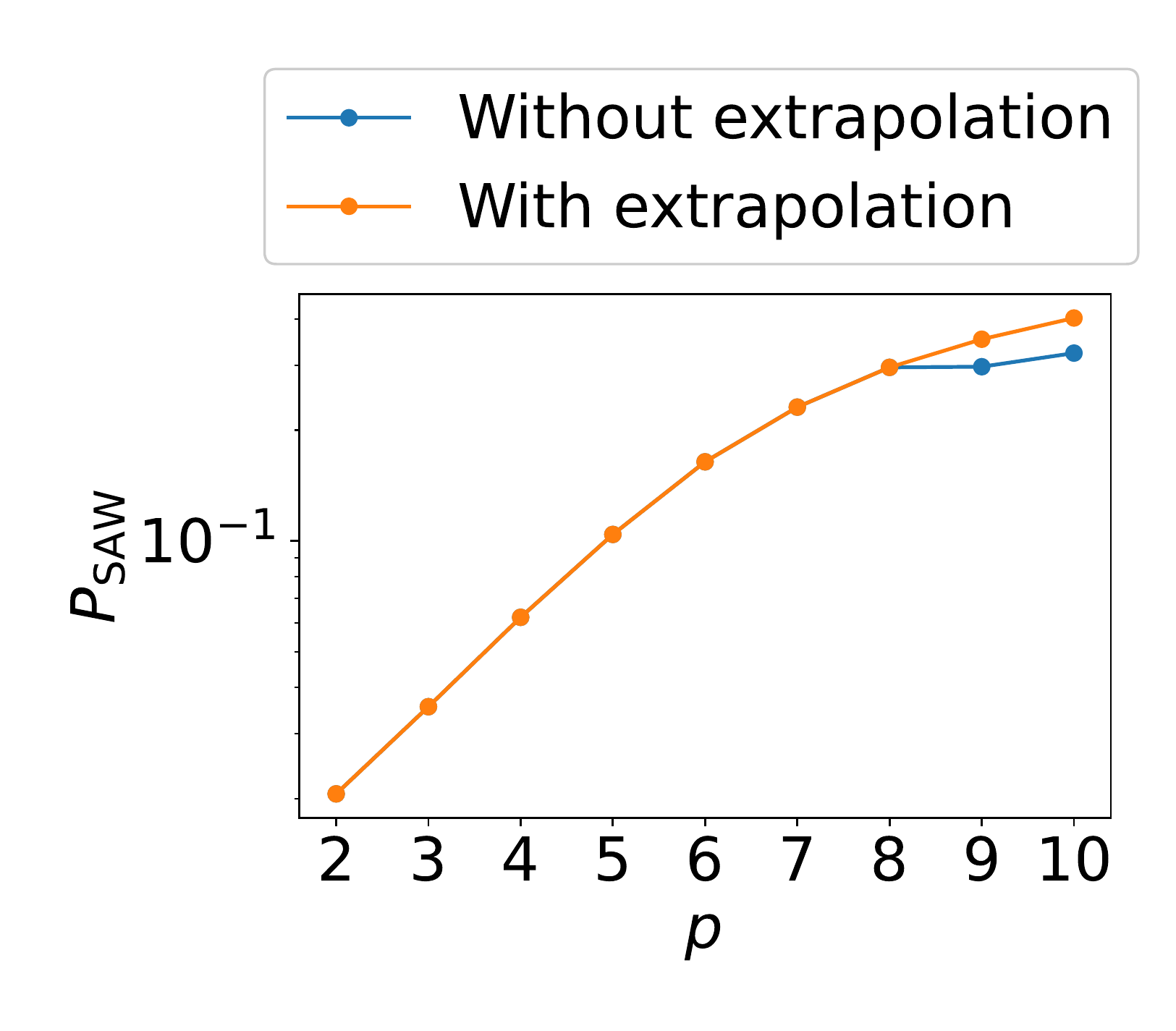}
        \caption{Success probability of QAOA optimized from random or extrapolated initialization}
        \label{subfig:with_without_extrapolation}
    \end{subfigure}
    \begin{subfigure}{0.32\textwidth}
        \centering
        \includegraphics[width=\textwidth]{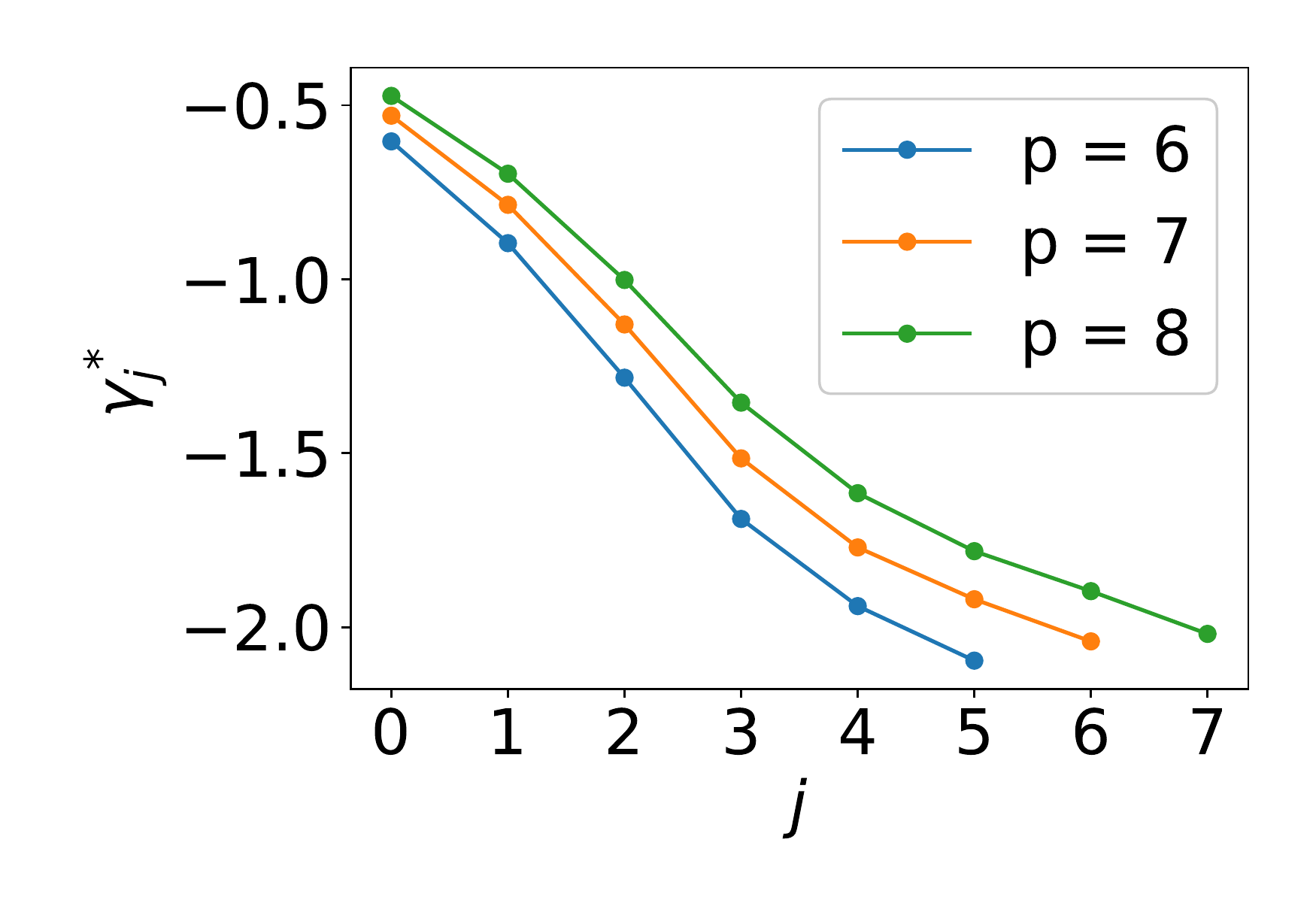}
        \caption{Optimal $\bm\gamma$ angles}
        \label{subfig:saw_gammas}
    \end{subfigure}
    \begin{subfigure}{0.32\textwidth}
        \centering
        \includegraphics[width=\textwidth]{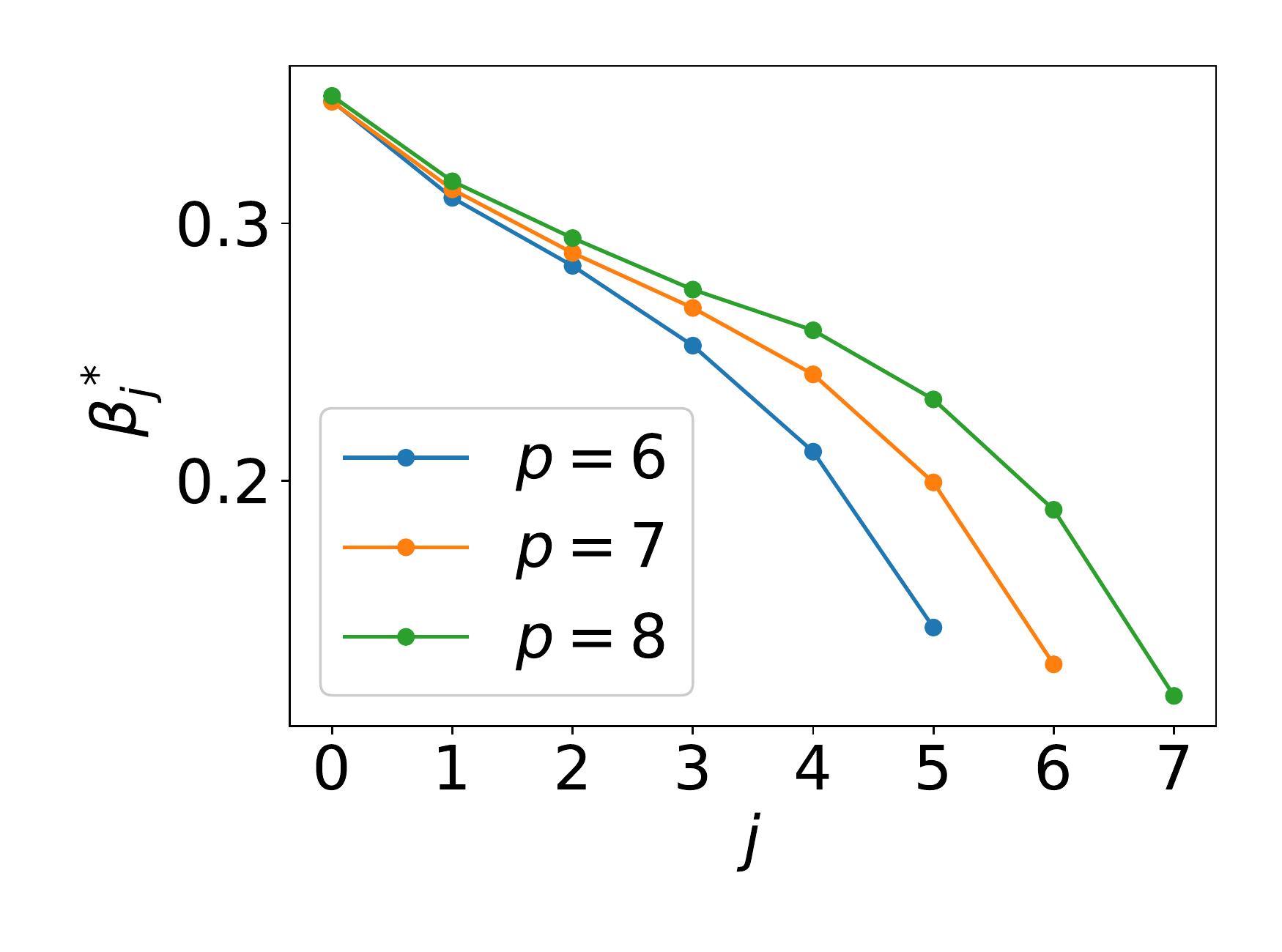}
        \caption{Optimal $\bm\beta$ angles}
        \label{subfig:saw_betas}
    \end{subfigure}
    \caption{Comparison of random and extrapolated variational parameters initialization (absolute turn-based encoding, inversion about the mean mixer)}
    \label{fig:saw_extrapolation}
\end{figure}

We conclude by comparing the random and extrapolation methods for initializing the ansatz. Figure \ref{subfig:with_without_extrapolation} compares the probability of sampling a self-avoiding walk for optimal variational parameters given by each method, using an encoding allowing backtracking and an ``inversion about the mean" mixer. Extrapolation eventually outperforms random initialization from $p = 9$, suggesting it should be preferred for optimizing the ansatz at large $p$. The validity of the approach can be justified by examining the optimal QAOA angles obtained from random initialization (figures \ref{subfig:saw_gammas} and \ref{subfig:saw_betas}), which appear to organize along a tractable pattern (monotonicity in angle index for a fixed ansatz depth, continuity in the ansatz depth).

\section{Sampling low-energy peptide conformations with QAOA}
\label{sec:sampling_low_energy_conformations}
Section \ref{sec:qaoa_saw} empirically showed that an efficiently trained, moderate-depth QAOA ansatz is capable of sampling a self-avoiding walk on a lattice with high probability. In this section, we turn to the complete peptide folding problem, where a folded peptide is modeled as a self-avoiding walk with attractive and repulsive interactions between its sites. We therefore investigate the capability of QAOA to sample a low-energy conformation among valid (self-avoiding) ones. The encoding of the problem (independent of the optimization algorithm) is described in section \ref{sec:encoding_generalities}; section \ref{sec:small_protein_qaoa_implementation} details the quantum algorithm used to address this problem; finally, section \ref{sec:ala_peptide_results} presents our results.

\label{sec:sampling_protein_conformations_qaoa}
\subsection{Encoding a conformation in qubits}
\subsubsection{Generalities}
\label{sec:encoding_generalities}

In the framework considered here, a protein conformation is sampled by preparing a quantum state and measuring it in the computational basis; the measured bitstring encodes a specific protein conformation. Several approaches exist to encode a protein conformation in a bitstring, see \cite{Babbush2014} for a review. Here, we use the  \textit{turn-based encoding}, introduced in this previous study and used in many subsequent works on quantum algorithms for protein folding \cite{PerdomoOrtiz2012,1810.13411,1811.00713,Robert2021,2101.10279}.

The turn-based encoding represents each chain (backbone and side chains) of the peptide as a sequence of ``turns". Precisely, each atom of the chain is described by its relative position with respect to the previous atom in the chain; for a lattice-based protein model, the allowed relative positions are finite and correspond to the basis vectors of the lattice. Each turn can then be digitally encoded into a finite number of bits, and the sequence of these turns (for all chains) is therefore represented as a bitstring.

In this work, we consider the lattice protein model described in \cite{Robert2021}, where atoms lie on a regular tetrahedral lattice. As discussed by the authors, such a choice may be justified by the interbond and dihedral angles commonly observed in physical conformations. However, we note that due to limited resources on quantum hardware or simulators, the previous work eventually modelled the protein as a sequence of amino acids and not at the finer-grained atomic level. Now, while the choice of a tetrahedral lattice favours realistic geometries when modelling the protein atom by atom, it is not obvious that this property persists when coarse-graining the protein at the amino acid level. In contrast, in this work, the conformation of the protein is described by the positions of all heavy atoms from the backbone chain. This has the advantage of facilitating comparison between the conformations generated by the quantum algorithms and the ones produced by classical methods such as molecular dynamics and metadynamics. On the flip side, this choice requires more encoded turns, hence more qubits, restricting us to smaller proteins than in \cite{Robert2021}. Besides, even for the trivial peptide size (2-4 residues) considered in this work, the limited number of qubits available on classical simulators of quantum computers restricts us to modelling heavy atoms from the backbone chain, factoring out lighter atoms (H) and all side chains (see section \ref{sec:hamiltonian_evolution} for details).

\subsubsection{Details}
\label{sec:encoding_details}
We now precisely describe the turn-based encoding used in this work. Our proposal is a slight modification from the one in \cite{Robert2021}, where the non-backtracking constraint is automatically enforced.

\begin{figure}[!tbp]
    \centering
    \begin{subfigure}{0.2\textwidth}
	    \centering
	    \includegraphics[width=0.8\textwidth]{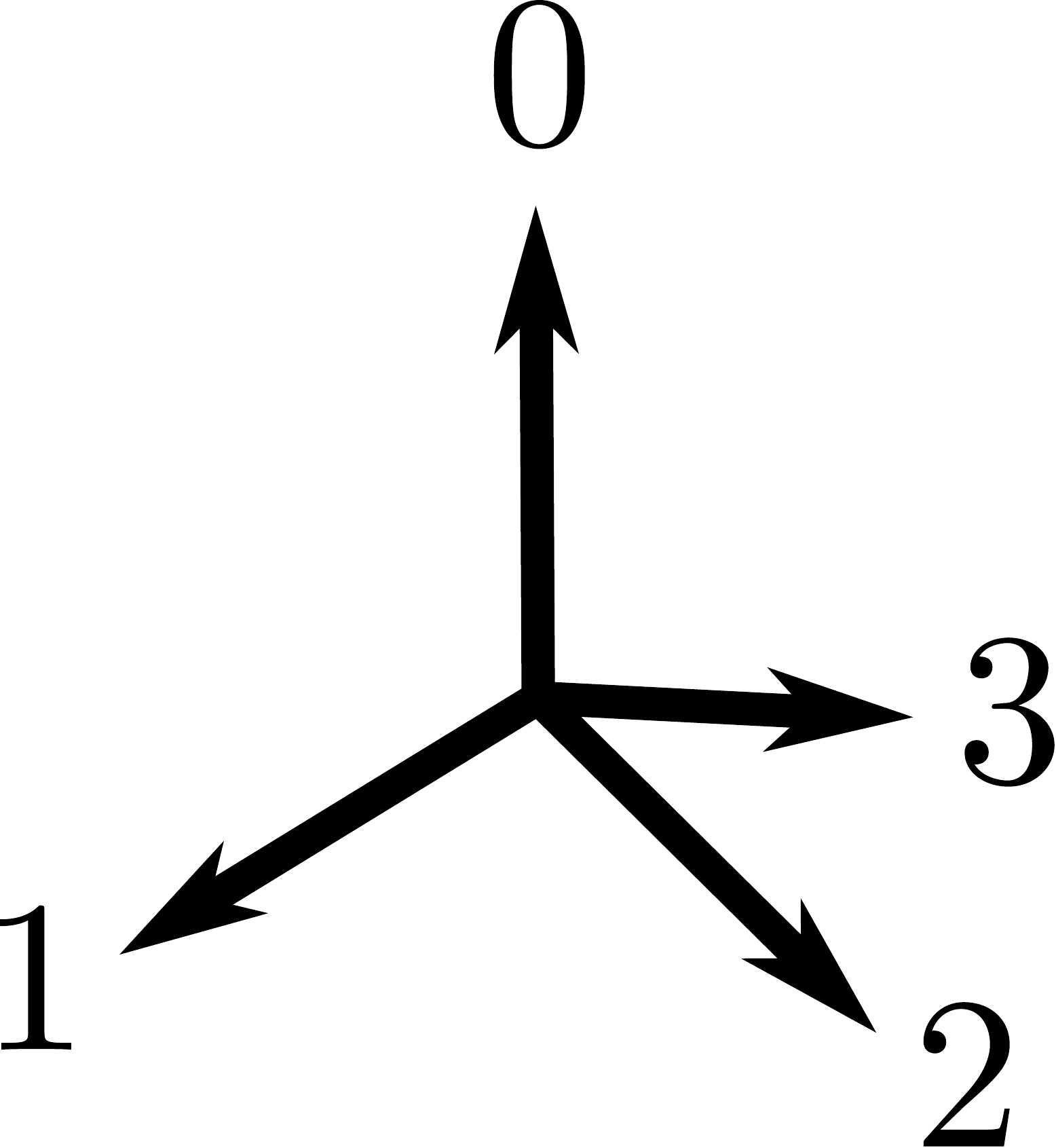}
	    \caption{Even turns}
    \end{subfigure}
    \begin{subfigure}{0.2\textwidth}
	    \centering
	    \includegraphics[width=0.8\textwidth]{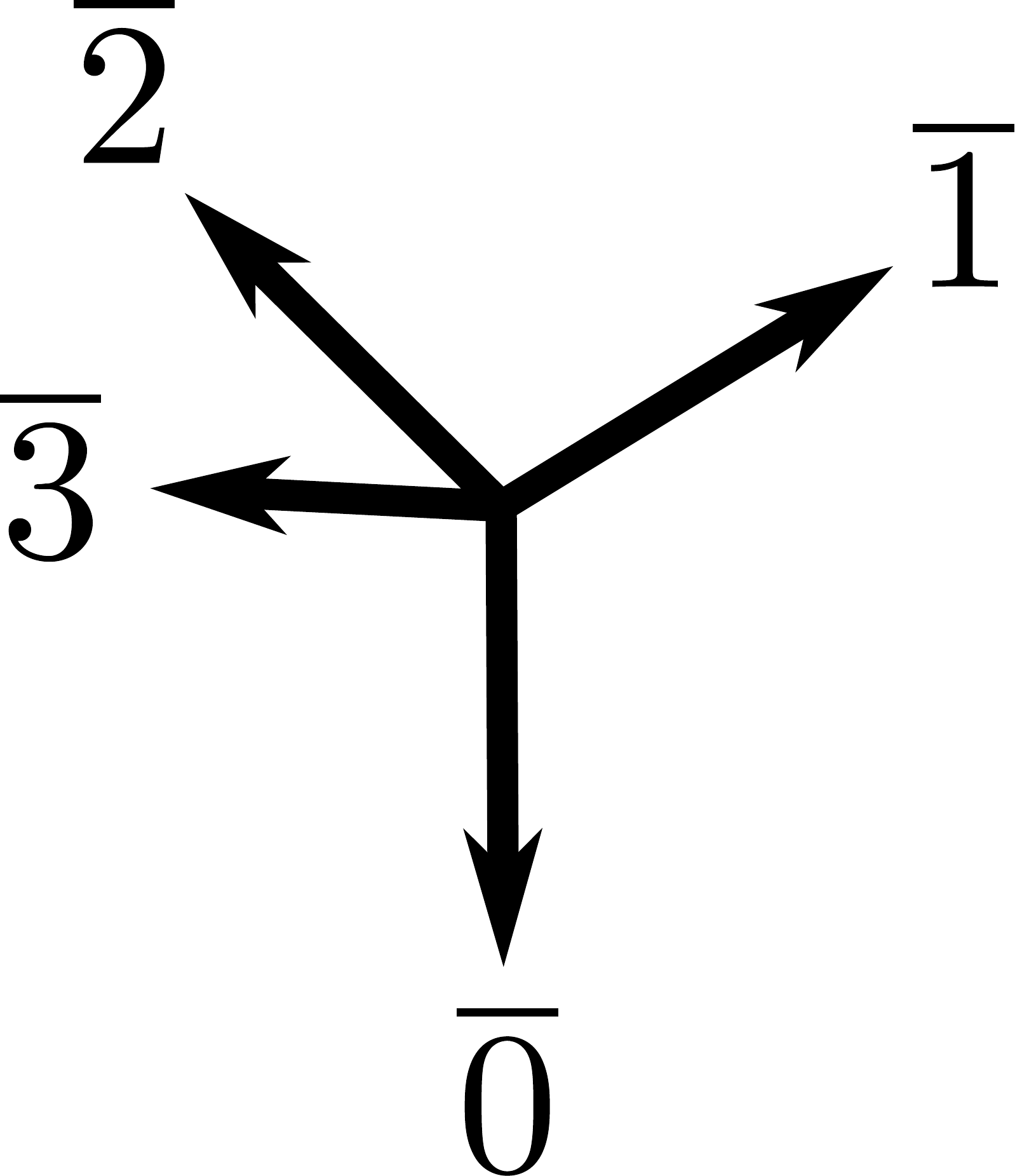}
	    \caption{Odd turns}
    \end{subfigure}
	\begin{subfigure}{0.5\textwidth}
		\centering
		\includegraphics[width=0.9\textwidth]{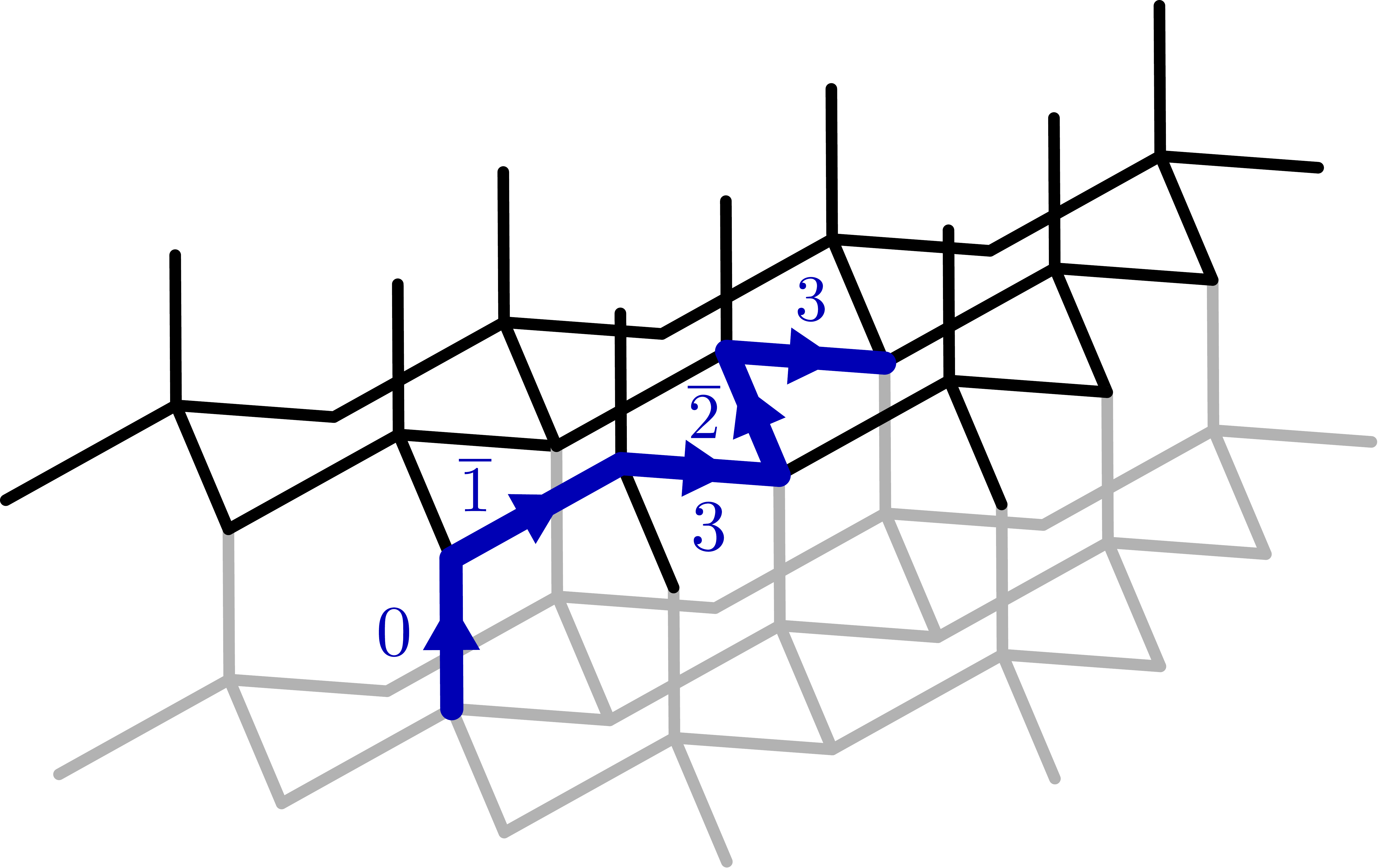}
		\caption{Walk on tetrahedral lattice}
	\end{subfigure}
    \caption{Turn encoding on a tetrahedral lattice}
\label{fig:turn_encoding}
\end{figure}

In this previous work, each turn on the tetrahedral lattice is represented by an integer $k \in \{0, 1, 2, 3\}$, which can be encoded using 2 (binary encoding) or 4 (unary encoding) bits. In this work, we retain the binary encoding to limit the number of required qubits; therefore, each turn is encoded in a ququart (consisting of two qubits). The interpretation of the integer as a turn direction depends on the parity of the turn index (figure \ref{fig:turn_encoding}). More precisely, for an even turn index, integer $k \in \{0, 1, 2, 3\}$ encodes a turn direction opposite to the direction it encodes for an odd turn index; following \cite{Robert2021}, the former and latter directions are respectively denoted by $k$ and $\overline{k}$ on the figure. A chain is then represented by a sequence of integers $k \in \{0, 1, 2, 3\}$ and the non-backtracking condition amounts to requiring that consecutive integers in the sequence be distinct. Besides, note, as discussed in \cite{Robert2021}, that one may without loss of generality fix the values of the first two turns thanks to the rotational symmetry of the tetrahedral lattice.

Here, we propose a slightly more economical\footnote{As measured by the dimension of the space of encoded configurations.} encoding, whereby each turn is encoded by an integer $k' \in \{0, 1, 2\}$. Given the value of the previous turn and the non-backtracking constraint, $k'$ indexes the allowed values for the current turn. We will say that $k'$ encodes the turn in a \textit{relative} way, whereas in the encoding described in the previous paragraph, $k$ encoded \textit{absolute} turns. The explicit mapping between the relative and absolute encoding of the turns is given in appendix \ref{sec:extra_figures}, table \ref{tab:relative_to_absolute_encoding}.

The relative turn encoding has the advantage of automatically enforcing the non-backtracking condition, removing the need for a backtracking penalty term in the Hamiltonian. 

\subsection{QAOA implementation}
\label{sec:small_protein_qaoa_implementation}
\subsubsection{Mixer layer}
The relative turn-based encoding described in section \ref{sec:encoding_details} encodes the backbone of the protein into a sequence of registers taking 3 possible values. These registers may then be regarded as qutrits. This suggests to use the mixer for QAOA on qudits proposed in \cite{2011.13420}. Applied to qutrits, this mixer depends on two angles $\beta_0, \beta_1$ and its action on $n - 3$ qutrits can be written as:
\begin{align}
    U_B\left(\beta_0, \beta_1\right) & := \left(e^{-\frac{i\beta_0}{2}}\ket{+}\bra{+} + e^{-\frac{i\beta_1}{2}}Z_3\ket{+}\bra{+}Z_3^{\dagger} + Z_3^2\ket{+}\bra{+}\left(Z_3^2\right)^{\dagger}\right)^{\otimes(n - 3)}\\
    & = \left(e^{-\frac{i\beta_0}{2}\ket{+}\bra{+} - \frac{i\beta_1}{2}Z_3\ket{+}\bra{+}Z_3^{\dagger}}\right)^{\otimes(n - 3)},
\end{align}
where
\begin{align}
\ket{+} & = \frac{\ket{0} + \ket{1} + \ket{2}}{\sqrt{3}}\\
    Z_3 & := \begin{pmatrix}
    1 & 0 & 0\\
    0 & \omega & 0\\
    0 & 0 & \omega^2
    \end{pmatrix}\\
    \omega & := e^{\frac{2\pi i}{3}}
\end{align}
In fact, it is often empirically sufficient (see section \ref{sec:qaoa_saw}) to let $\beta_1 = 0$ in the formula above, with $\beta_1 \neq 0$ achieving at best a marginal improvement while making the variational optimization considerably more difficult. In this case, the mixer reduces to:
\begin{align}
    U_B(\beta_0, 0) & = \left(e^{-\frac{i\beta_0}{2}\ket{+}\bra{+}}\right)^{\otimes (n - 3)}\\
    & = \left\{\mathbf{1}_{3 \times 3} + \left(e^{-\frac{i\beta_0}{2}} - 1\right)\ket{+}\bra{+}\right\}^{\otimes(n - 3)}.
\end{align}
The case $\beta_0 = 2\pi$ corresponds to applying an inversion about the mean operator $\mathbf{1}_{3 \times 3} - 2\ket{+}\bra{+}$ to each qutrit; therefore, we will occasionally refer to this mixer as an \textit{``inversion about the mean" mixer}. Note that this differs from the Grover mixers introduced in \cite{bartschi_2020}, where the inversion about the mean acts on all qubits and not independently on each qudit as here.

\subsubsection{Problem Hamiltonian}
\label{sec:hamiltonian_evolution}

Having described the problem's encoding and the corresponding choice of mixer, it remains to specify the problem's cost function. In molecular dynamics, a potential energy function describing the energy of a molecule is called a \textit{force field}. A common choice is the CHARMM force field and its variants \cite{Reiher,MacKerell1998,Mackerell2004}. This empirically fitted potential depends on many degrees of freedom, including the bond lengths, interbond angles and dihedral angles and can be calculated on a quantum computer using quantum arithmetic. Detailed estimates of the needed quantum resources were derived in \cite{2105.09690} based on the systematic analysis of quantum arithmetic circuits~\cite{1805.12445}. The authors calculated that for an $N$-atom protein, each of which has Cartesian coordinates encoded on $b$ qubits, either $19b$ ancillary qubits and a Toffoli depth $\frac{52N(N - 1)}{2}$, or $\frac{19bN}{2}$ ancillary qubits and a Toffoli depth $51(N - 1)$ were required to evaluate the most costly contribution of the force field.

In this work, the highly constrained geometry of discretized protein conformations does not justify using the complex force field just described. We therefore focus on the computationally most expensive part of the force field: the Lennard-Jones potential. Specifically, we resort to an economically parametrized Lennard-Jones potential as proposed in \cite{Schauperl2020}. The Lennard-Jones potential can be expressed as a sum of two-body interactions between all pairs of atoms:
\begin{align}
    H_{\textrm{Lennard-Jones}} & := \sum_{\textrm{atom pairs} \{i, j\}}H_{\textrm{Lennard-Jones}, \{i, j\}},\\
    H_{\textrm{Lennard-Jones}, \{i, j\}} & = \sqrt{\varepsilon_i\varepsilon_j}\left(\left(\frac{r_{1/2, i} + r_{1/2, j}}{\left|\bm r_i - \bm r_j\right|}\right)^{12} - 2\left(\frac{r_{1/2, i} + r_{1/2, j}}{\left|\bm r_i - \bm r_j\right|}\right)^6\right)
\end{align}
where $\bm r_i$ is the position of atom $i$ and the $r_{1/2, i}, \varepsilon_i$ are parameters specific to each atom. In the simplest model described in \cite{Schauperl2020}: the HCON model adopted here, $r_{1/2, i}$ and $\varepsilon_i$ only depend on the nature of atom $i$ (hydrogen, carbon, oxygen, nitrogen). Despite its simplicity, this model is reported \cite{Schauperl2020} to yield quantitatively accurate molecular dynamics simulations.

In this study, we apply the Lennard-Jones model to short sequences of alanine amino acids, which are a common benchmark for molecular dynamics simulations (see e.g.~\cite{Hu2003}); the alanine tetrapeptide considered in the rest of this section is shown in figure \ref{fig:ala_tetrapeptide} from appendix \ref{sec:extra_figures} for reference.
More precisely, the total energy attributed to a conformation comprises a Lennard-Jones contribution and a penalization of clashes (atoms occupying the same sites):
\begin{align}
\label{eq:ala_peptide_full_potential}
    H & = H_{\textrm{Lennard-Jones}} + \lambda \times \left(\textrm{number of clashes}\right),
\end{align}
where $\lambda > 0$ is a tunable penalty coefficient. In the equation above, it is implicitly understood that the $H_{\textrm{Lennard-Jones}, \{i, j\}} = 0$ for a pair $\{i, j\}$ of overlapping atoms. The difficulty is that the potential \ref{eq:ala_peptide_full_potential} must be computed accounting for all atoms from the backbone and side chains, while the limited number of qubits available on classical simulators of quantum computers restricts us to encoding the positions of the heavy atoms (C, N, O) from the backbone chain. We address this problem by resorting to a partial minimization: to each encoded configuration of the heavy atoms from the backbone chain, we associate the full configuration with the compatible backbone chain of lowest energy; we then attribute this energy to the encoded conformation. In optimizing over positions of atoms side chain atoms, we restricted these to live on the lattice, though this is not strictly necessary. Finally, atomic bond distances were always fixed to $1.5\,\mathring{\mathrm{A}}$. This corresponds in order of magnitude to the bond distances observed in the alanine amino acid of the simple benchmark pentapeptide introduced in \cite{Scherer2015}. We underline that the sole purpose of this definition is to make the study of quantum optimization algorithms tractable with a limited number of qubits (for instance, a tetrapeptide can be encoded using 20 qubits, a number where variational optimization of quantum circuits with many layers remains feasible in classical emulation). In particular, this approach does not scale up, since the number of discrete variables on which the partial minimization is carried out (degrees of freedom of side chains) grows linearly with the number of amino acids; this implies, \textit{a priori}, an exponential scaling of the depth of a quantum circuit realizing this partial minimization. Besides, it is theoretically unclear, both in general and in this particular case, whether partial minimization of the cost function degrades or improves the performance of QAOA.

\subsubsection{Optimization of variational parameters}
\label{sec:ala_peptide_variational_optimization}

Optimizing the variational parameters of the QAOA ansatz is known to be a hard problem both computationally \cite{2101.07267} and in practice \cite{Zhou2020} and has been the subject of various theoretical \cite{hogg00,Wang2018,2103.11976} and numerical \cite{1812.04170,Zhou2020,Sack2021,pmlr-v107-yao20a,Khairy2020,1911.04574,Johnson2011,2110.10685} studies. In particular, finding good parameter initialization strategies is crucial for optimizing the ansatz, without which a number of optimization attempts exponential in the number of layers may be required to reach the global minimum \cite{Zhou2020}.

In this study, 3 parameter initialization strategies were compared:
\begin{itemize}
    \item \textbf{Random initialization}: all angles $\beta_j, \gamma_j$ are drawn uniformly at random.
    \item \textbf{Quantum annealing schedule}: inspired by \cite{Sack2021}, this method initializes angles $\bm\beta, \bm\gamma$ according to a linear schedule mimicking a first-order Trotterization of quantum annealing for a time $\Delta t$: $\beta_j := -\frac{p - j}{p}\Delta t$, $\gamma_j := \frac{j + 1}{p}\Delta t$ for $j \in [p]$. The single parameter $\Delta t$ is optimized to minimize the expected energy.
    \item \textbf{Quantum annealing initialization}: inspired by \cite{Sack2021}, this technique uses the linear schedule previously described as the starting point of an unconstrained optimization.
    
\end{itemize}
All three methods can be combined with variational parameters extrapolation as described in the case of the self-avoiding walk in section \ref{sec:saw_experiments}. To facilitate the classical optimization of variational parameters $\bm\beta, \bm\gamma$ by gradient descent, the latters were rescaled so that the cost function assumes comparable gradients along all directions. This was practically done by visual inspection of the optimization landscape of the $p = 1$ QAOA; besides the rescaling, the angles were restricted to encompass the local minimum of the QAOA energy closest to $(\beta_0, \gamma_0) = (0, 0)$. Note there is no guarantee that this is the global minimum ---in fact, we conjecture it is not. An illustration is given on figure \ref{fig:qaoa_landscape_p=1} in appendix \ref{appendix:figures_small_protein}. This scaling and domain restriction are then generalized to higher QAOA levels $p$. Generalizing the rescaling is justified if one assumes optimal QAOA angles at level $p$ (in the restricted domain) to be dominated by optimal angles at level $p - 1$ (in the restricted domain), as verified in the case of the self-avoiding walk problem in section \ref{sec:qaoa_saw} (figures \ref{subfig:saw_betas} and \ref{subfig:saw_gammas}).

Whenever random initial parameters are required ($2p$ angles for random initialization or single annealing time $\Delta t$ for optimizing annealing schedule), $50$ initialization attempts were made to select the best result. Besides, the penalty coefficient $\lambda$ in equation \ref{eq:ala_peptide_full_potential} was set to $1000$; this is, in order of magnitude, the value beyond which the expected energy achieved by QAOA conditioned on the absence of clash (see results section \ref{sec:ala_peptide_results} and particularly figure \ref{fig:alanine_peptide_energy_optimization}) starts to degrade.

\subsection{Results}
\label{sec:ala_peptide_results}
In this section, we detail the results of training a QAOA ansatz to minimize the scoring function (equation \ref{eq:ala_peptide_full_potential}) of an alanine peptide. The discussion is restricted to the largest problem instance considered (tetrapeptide). All numerical experiments use the relative turn-based encoding introduced in section \ref{sec:encoding_details}.

\begin{figure}[!tbp]
	\centering
	\begin{subfigure}{0.45\textwidth}
		\centering
		\includegraphics[width=\textwidth]{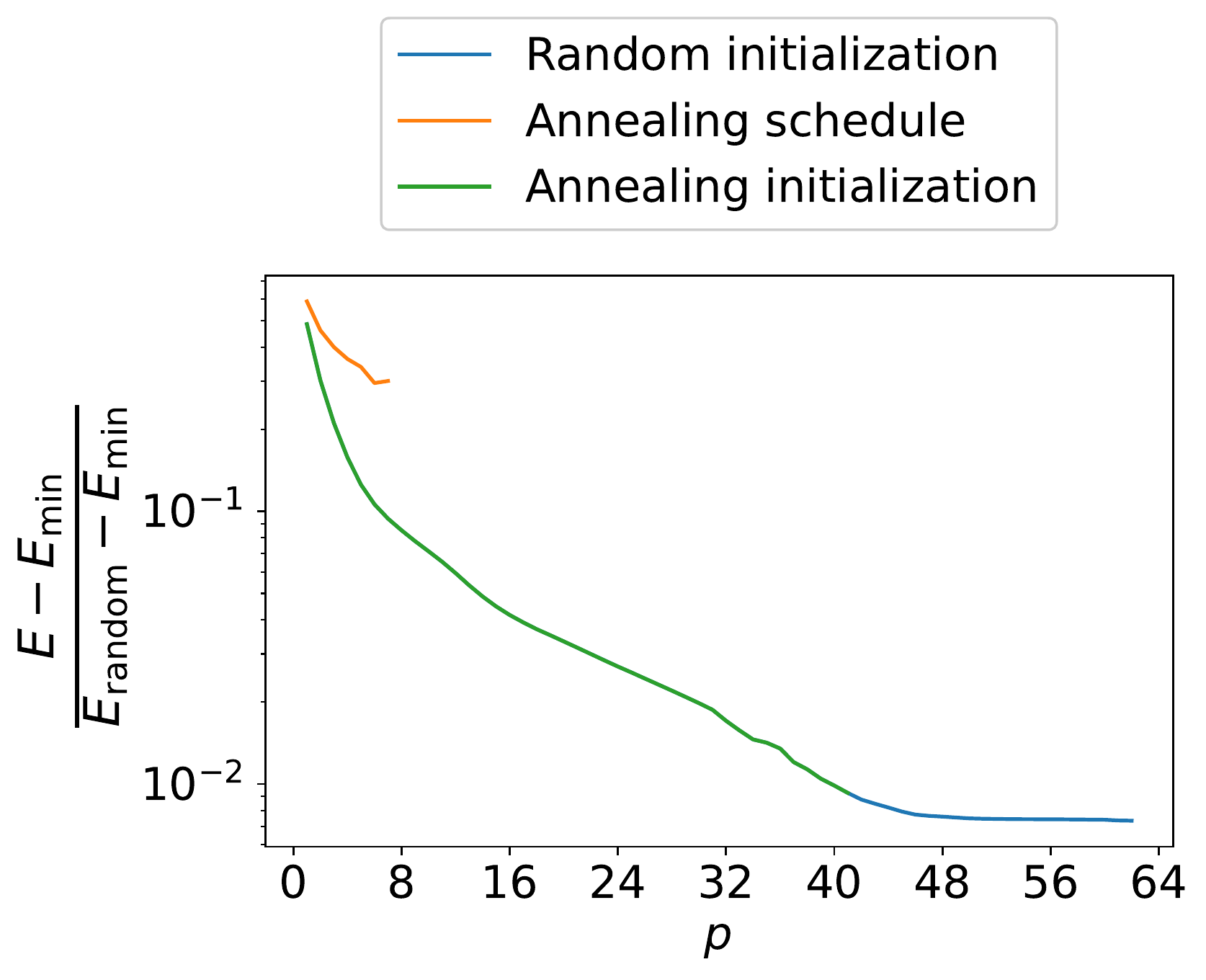}
		\caption{Total energy}
		\label{subfig:alanine_peptide_energy_optimization_energy}
	\end{subfigure}
	\hspace*{0.05\textwidth}
	\begin{subfigure}{0.45\textwidth}
		\centering
		\includegraphics[width=\textwidth]{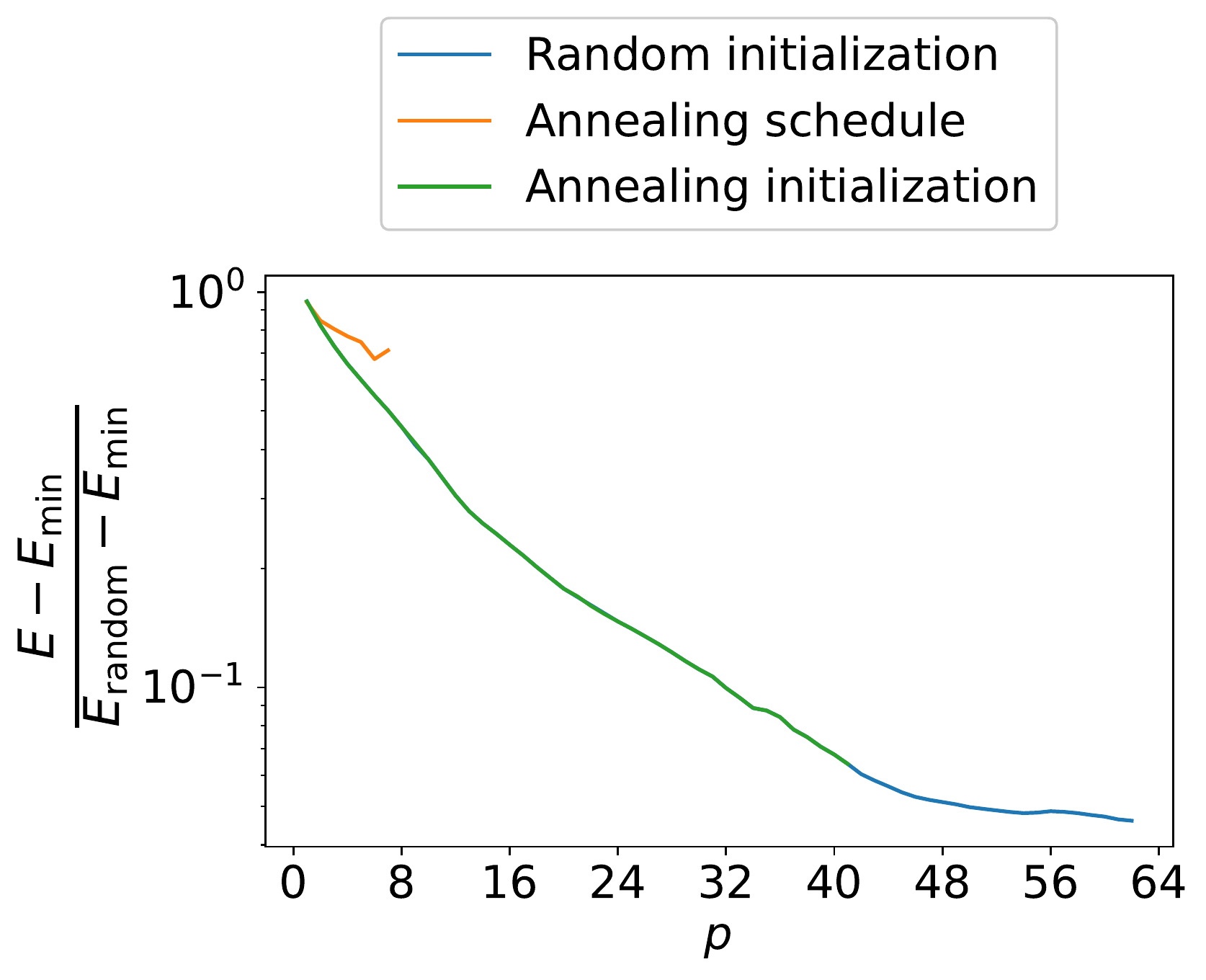}
		\caption{Energy given absence of clashes}
		\label{subfig:alanine_peptide_energy_optimization_energy_no_clash}
	\end{subfigure}
	\caption{Expected conformation energy achieved by QAOA for increasing ansatz depth and different angle optimization strategies. For random and annealing initializations, large $p$ parameters (random initialization: $p \geq 15$; annealing initialization: $p \geq 48$) were exclusively obtained by initializing the optimizer with variational angles extrapolated from lower $p$ angles. For both methods, extrapolation was performed from $p = 5$ and the best angles were retained between the extrapolated and non-extrapolated results.}
	\label{fig:alanine_peptide_energy_optimization}
\end{figure}

The expected energies achieved by the QAOA ansatz after training with different optimization strategies (random initialization, annealing schedule, optimization from optimized annealing schedule) are given on figure \ref{fig:alanine_peptide_energy_optimization}. The energy is expressed in a dimensionless form, as the distance to the minimum energy $E - E_{\mathrm{min}}$, rescaled by the same quantity evaluated at the energy given by random assignment (formally corresponding to $p = 0$ QAOA): $E_{\mathrm{random}} - E_{\mathrm{min}}$. This dimensionless energy varies from $1$ to $0$ as $E$ decreases from $E_{\mathrm{random}}$ to $E_{\mathrm{min}}$. In the previous expressions, $E$ is the expected energy of a conformation sampled from the variational circuit, $E_{\mathrm{random}}$ is the expected energy of a conformation sampled uniformly at random and $E_{\mathrm{min}}$ is the lowest possible energy of a conformation. We distinguish the cases where expectations are calculated over all conformations (figure \ref{subfig:alanine_peptide_energy_optimization_energy}) or over conformations without clashes only (figure \ref{subfig:alanine_peptide_energy_optimization_energy_no_clash}). In the former case, $E_{\mathrm{random}}$ is the expected energy of a uniformly sampled conformation (possibly with clashes) and $E$ is the expected energy of a conformation (possibly with clashes) sampled from the QAOA state; whereas in the latter case, $E_{\mathrm{random}}$ is the expected energy of a conformation sampled uniformly from conformations without clashes and $E$ is the expected energy of a conformation sampled from the QAOA state given it has no clash. 

\begin{figure}[!tbp]
    \centering
    \begin{subfigure}{\textwidth}
        \centering
        \begin{tabular}{c|c|c}
            \thead{Conformation} & \thead{Probability} & \thead{Energy ($\textrm{kcal.mol}^{-1}$)}\\
            \hline
            \adjustbox{valign=c}{\includegraphics[angle=90,width=0.3\textwidth]{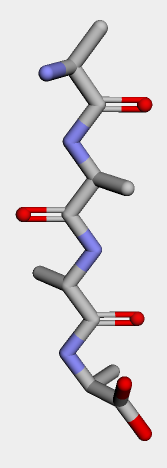}} & $2.33\cdot 10^{-3}$ & $5.23$\\
            \hline
            \adjustbox{valign=c}{\includegraphics[angle=90,width=0.3\textwidth]{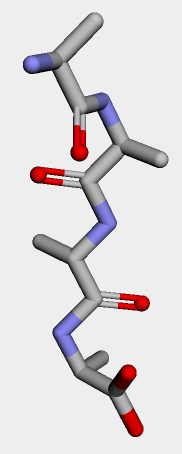}} & $2.19\cdot 10^{-3}$ & $5.05$\\
            \hline
            \adjustbox{valign=c}{\includegraphics[angle=90,width=0.3\textwidth]{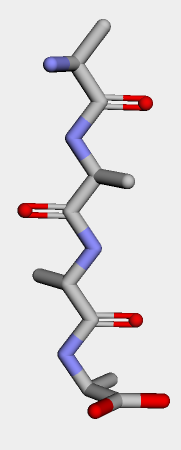}} & $2.15\cdot 10^{-3}$ & $5.03$
        \end{tabular}
        \caption{Most frequent conformations}
    \end{subfigure}\\
    \vspace*{20px}
    \begin{subfigure}{0.4\textwidth}
        \centering
        \hspace*{-30px}
        \begin{tabular}{c|c|c}
            \thead{Conformation} & \thead{Probability} & \thead{Number of\\ clashes}\\
            \hline
            \adjustbox{valign=c}{\includegraphics[width=0.5\textwidth]{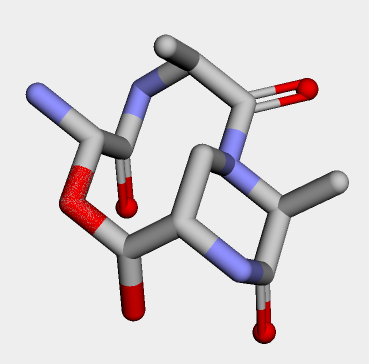}} & $1.15\cdot 10^{-6}$ & $6$\\
            \hline
            \adjustbox{valign=c}{\includegraphics[width=0.5\textwidth]{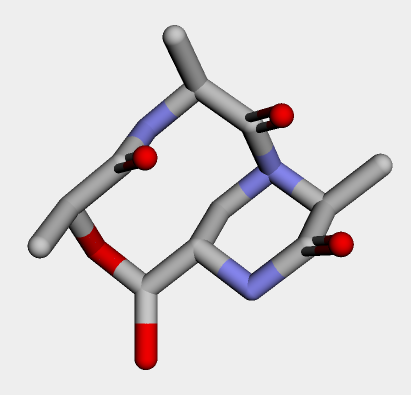}} & $7.28\cdot 10^{-7}$ & $2$\\
            \hline
            \adjustbox{valign=c}{\includegraphics[width=0.5\textwidth]{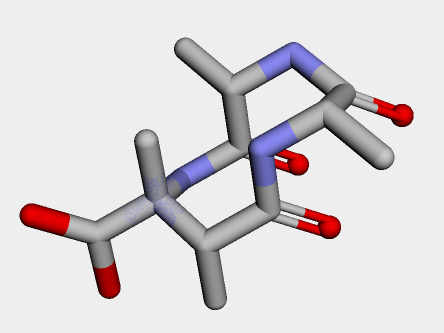}} & $3.26\cdot 10^{-7}$ & $5$
        \end{tabular}
        \caption{Most frequent invalid conformations}
    \end{subfigure}
    \hspace*{0.02\textwidth}
    \begin{subfigure}{0.45\textwidth}
        \centering
        \begin{tabular}{c|c|c}
            \thead{Conformation} & \thead{Probability} & \thead{Energy\\ ($\textrm{kcal.mol}^{-1}$)}\\
            \hline
            \adjustbox{valign=c}{\includegraphics[width=0.45\textwidth]{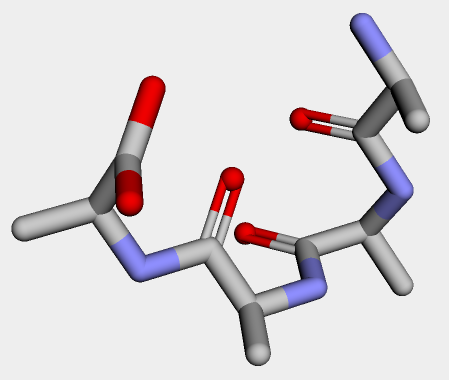}} & $1.22\cdot 10^{-4}$ & $-5.64$\\
            \hline
            \adjustbox{valign=c}{\includegraphics[width=0.45\textwidth]{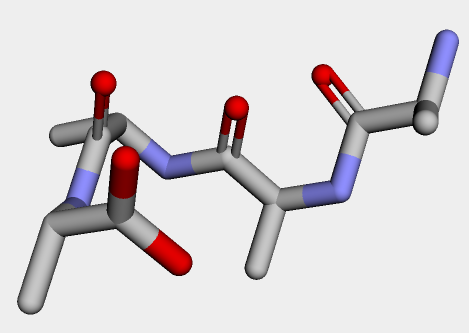}} & $1.38\cdot 10^{-4}$ & $-5.54$\\
            \hline
            \adjustbox{valign=c}{\includegraphics[width=0.45\textwidth]{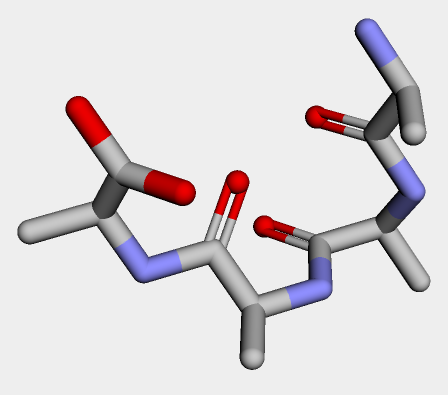}} & $1.22\cdot 10^{-4}$ & $-5.49$
        \end{tabular}
        \caption{Lowest-energy conformations sampled by the algorithm}
    \end{subfigure}
    \caption{Sample conformations generated by the algorithm (reported energies simply account for the simple Lennard-Jones potential used in this work and may differ from predictions by more sophisticated methods). Clashes refer to atoms (including $H$) from either the backbone or side chains occupying the same tetrahedral lattice site as detailed in section \ref{sec:hamiltonian_evolution}.}
    \label{fig:conformation_examples}
\end{figure}

These results first show that constraining variational parameters to follow an annealing schedule leads to a highly suboptimal expected energy, but the latter is considerably improved after optimizing all angles starting from this annealing schedule. However, random initialization combined with extrapolation from a modest $p=5$ ultimately seems to outperform these techniques. Numbers supporting these statements are provided in the captions of the figures. Around level $p = 100$, the relative improvement over random assignment is of order $10^{-1}$, meaning $90\,\%$ of the achievable improvement over random has been achieved. Note that the improvement of the expected total energy (Lennard-Jones potential and clash penalization) over random assignment is more important than the improvement of the expected energy in absence of clashes (Lennard-Jones contribution only). We attribute this to the efficiency of the ansatz at suppressing clashes; this is illustrated by further numerical results in appendix \ref{appendix:figures_small_protein}. Finally, similar to the self-avoiding walk study, we explicitly illustrated the merit of variational parameters extrapolation; the numerical results are reported in the same appendix. As a more concrete representation of our results, we report in figure \ref{fig:conformation_examples} the most frequent valid or invalid conformations sampled by the QAOA ansatz. We also show the lowest energy conformations and their probabilities of being sampled; the discretized conformation space has exactly $59049$ conformations.

\begin{figure}[!tbp]
	\centering
	\begin{subfigure}{0.75\textwidth}
		\centering
		\includegraphics[width=\textwidth]{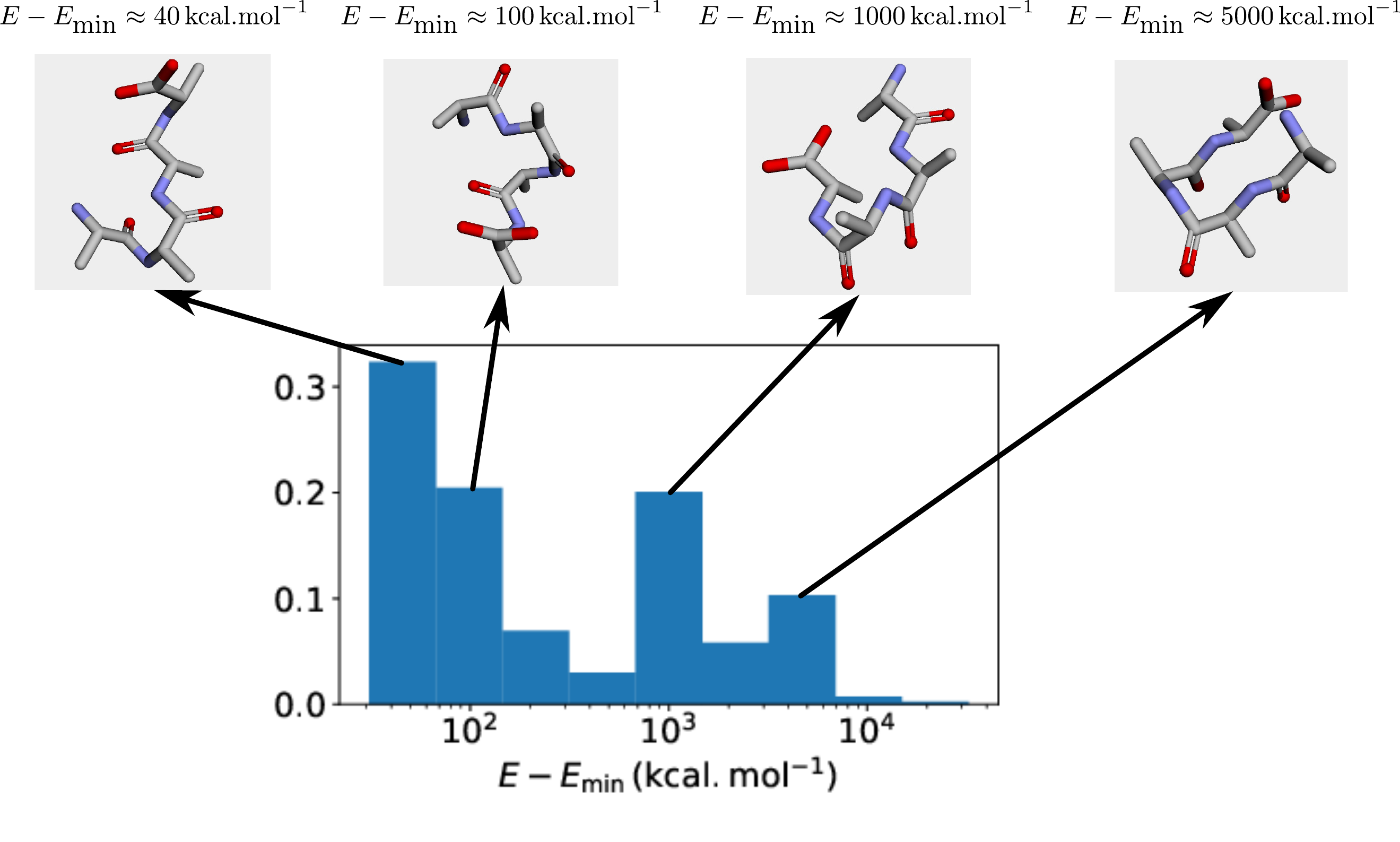}
		\caption{$p = 0$ (random assignment)}
	\end{subfigure}\\
	\begin{subfigure}{0.32\textwidth}
		\centering
		\includegraphics[width=\textwidth]{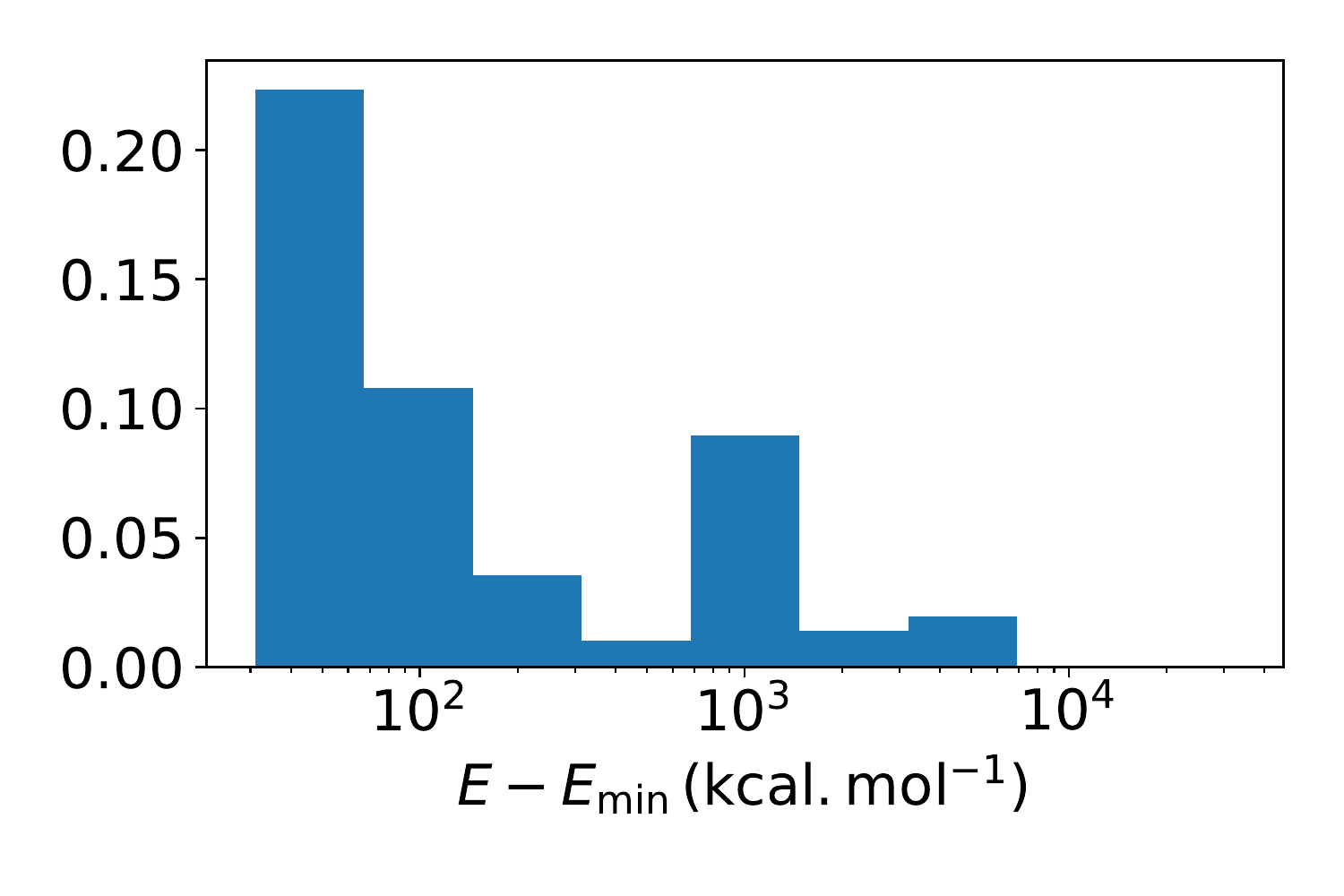}
		\caption{$p = 8$}
	\end{subfigure}
	\begin{subfigure}{0.32\textwidth}
		\centering
		\includegraphics[width=\textwidth]{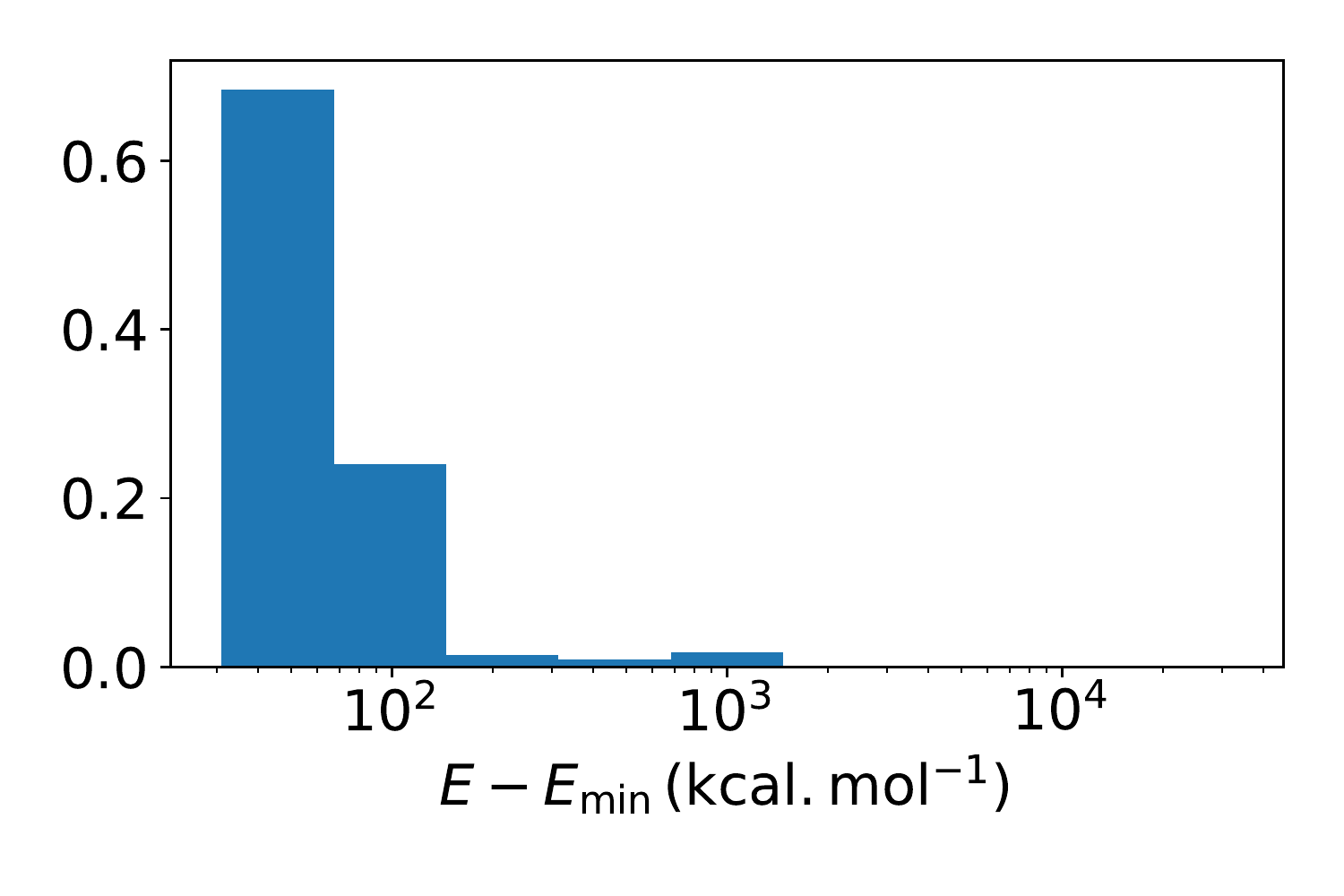}
		\caption{$p = 62$}
	\end{subfigure}
	\caption{Energy distribution of conformations sampled from QAOA ansatz for $p \in \{0, 8, 62\}$ ($p = 0$ corresponds to a uniformly random conformation); $E_{\textrm{min}} = -36.69\,\textrm{kcal.mol}^{-1}$ is a lower bound for the Lennard-Jones potential obtained by minimizing all pairwise interactions independently (in particular, all physical conformations have a strictly higher energy). Random initialization strategy for variational parameters. For $p = 0$, example conformations from different areas of the histogram are represented.}
	\label{fig:ala_peptide_energy_histogram}
\end{figure}

\begin{figure}[!tbp]
    \centering
    \begin{subfigure}{0.32\textwidth}
        \centering
        \includegraphics[width=\textwidth]{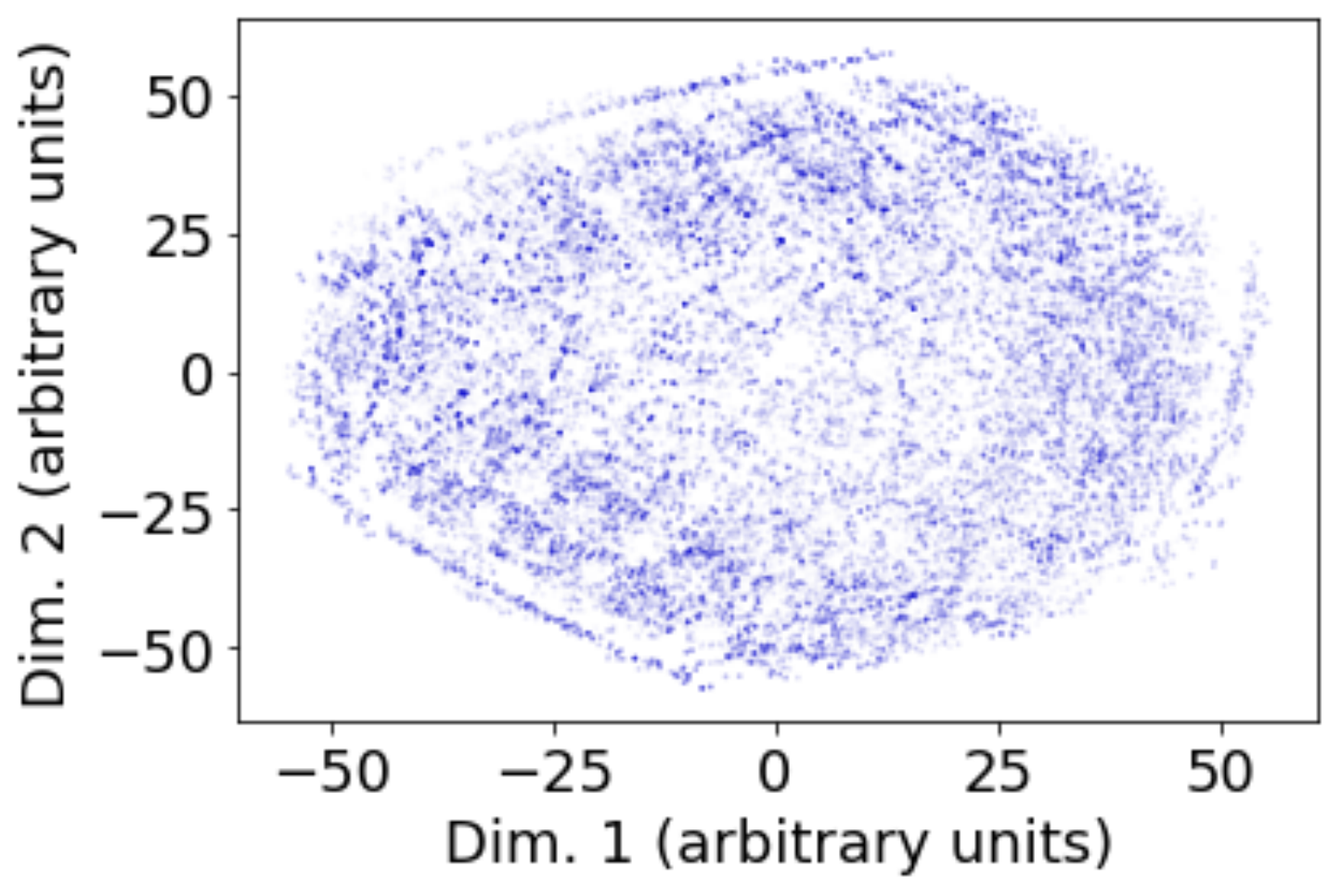}
        \caption{$p = 2$}
    \end{subfigure}
    \begin{subfigure}{0.32\textwidth}
        \centering
        \includegraphics[width=\textwidth]{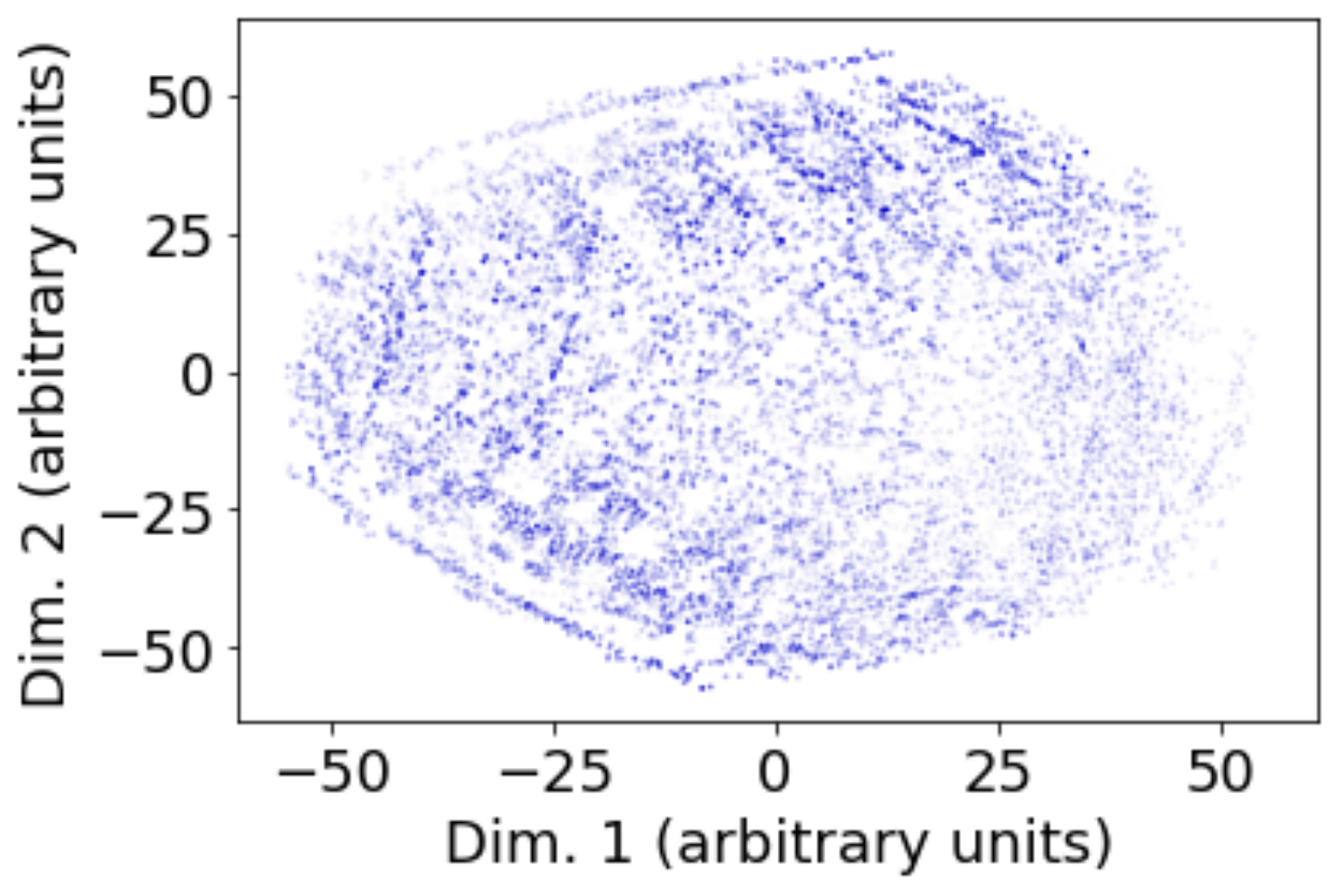}
        \caption{$p = 8$}
    \end{subfigure}
    \begin{subfigure}{0.32\textwidth}
        \centering
        \includegraphics[width=\textwidth]{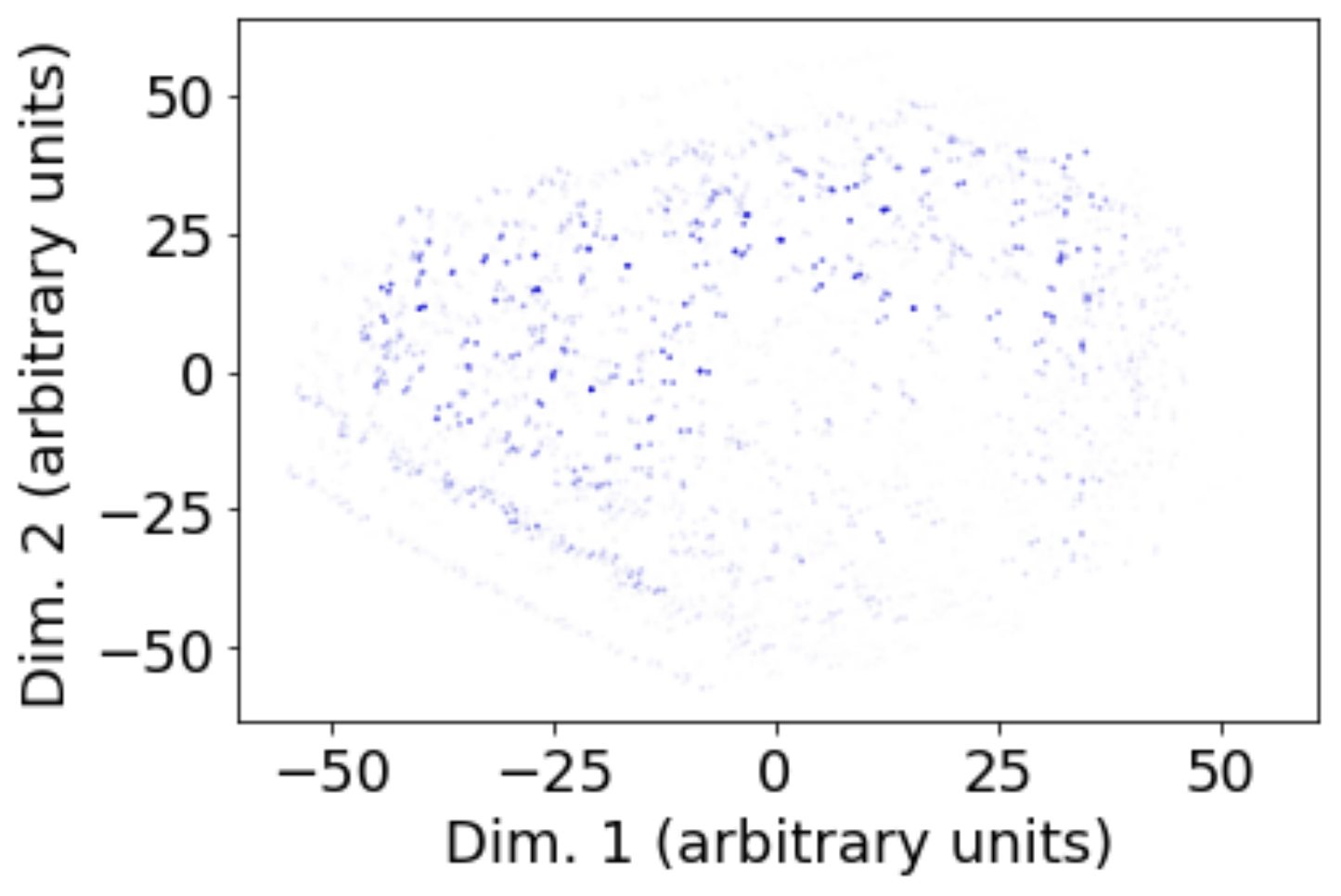}
        \caption{$p = 62$}
    \end{subfigure}
    \caption{Multidimensional scaling for alanine tetrapeptide, with random initialization and extrapolation. Probabilities are represented on a relative alpha scale ($\alpha = 0$: $0$ probability 0, $\alpha = 1$: maximum probability).}
    \label{fig:conformations_mds}
\end{figure}

While the energy in equation \ref{eq:ala_peptide_full_potential} is the cost function used to build and optimize the QAOA ansatz, other metrics can be used to evaluate conformations sampled from the ansatz once trained. A possibility is to consider the full energy distribution of the sampled conformations. The latter is represented as a histogram in figure \ref{fig:ala_peptide_energy_histogram} for different ansatz depths $p$.
Besides, we give (figure \ref{fig:conformations_mds}) a representation of this distribution, known as \textit{multidimensional scaling}, that accounts for the geometric distance between the configurations. These configurations, initially represented by 3 coordinates for each atom, are mapped to two-dimensional vectors whose mutual distances attempt to reproduce those of the original vectors; we refer e.g. to \cite{mds_2008} for technical details. The axis of symmetry in the 2D representation is related to the invariance of the energy under reflection about the plane defined by the initial two turns. The general hexagonal (instead of circular) shape of the points can be attributed to the non-isotropy of the tetrahedral lattice. Finally, the non-uniformity of the density of points can be explained by the non-independence of the components of the vectors representing conformations, which must be self-avoiding walks.

\begin{figure}[!tbp]
	\centering
	\includegraphics[width=0.65\textwidth]{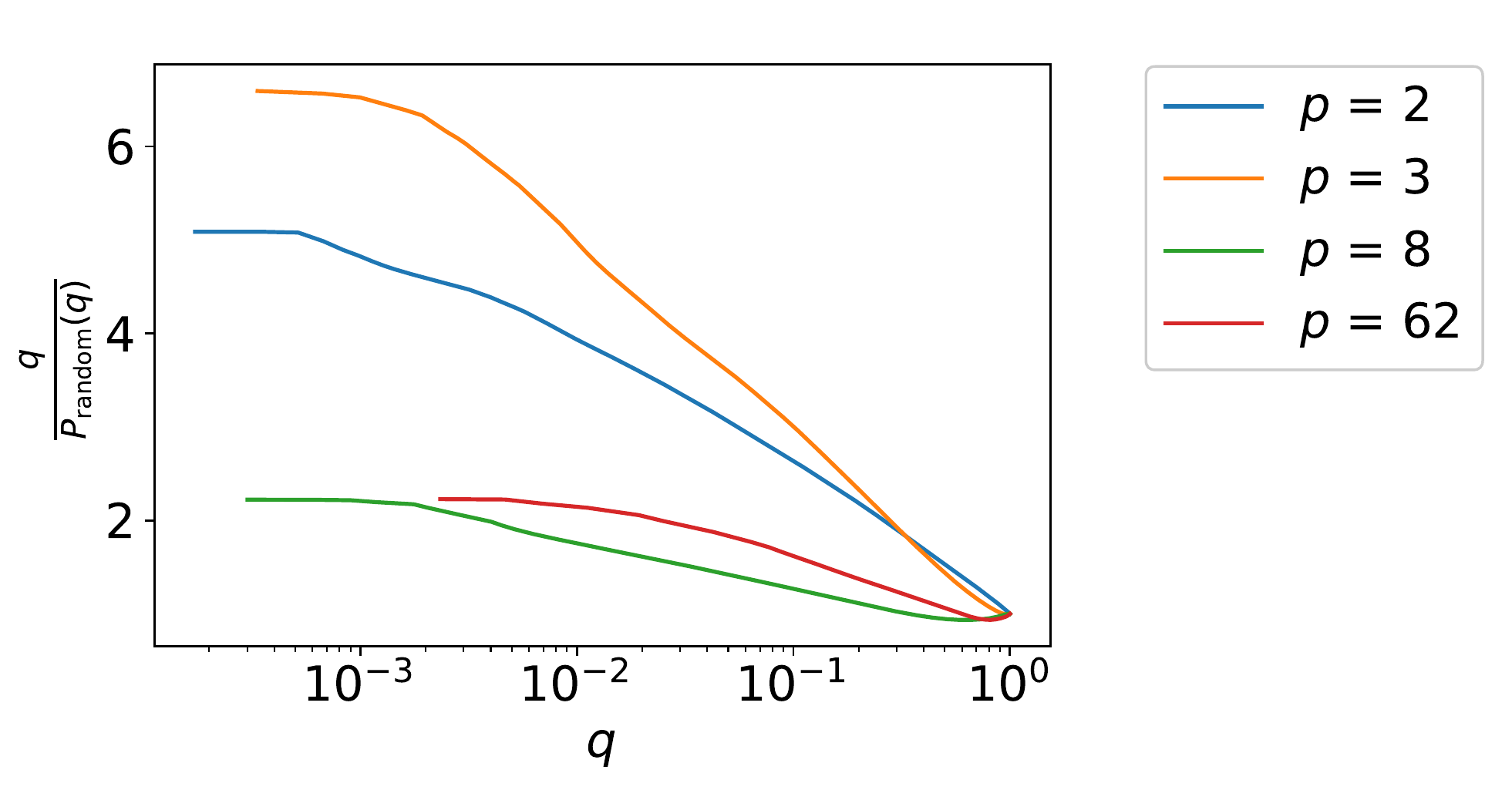}
	\caption{Comparison of QAOA and random guessing for sampling a conformation in the $q$ quantile of the QAOA distribution. The represented metric is essentially the ratio between the number of random guessing attempts and the number of QAOA sampling attempts to obtain a conformation pertaining to a given quantile of the QAOA distribution.}
	\label{fig:ala_peptide_qaoa_vs_random_guessing}
\end{figure}

Several summary metrics may be extracted from these distributions. We propose a specific figure of merit that allows for a fair comparison between QAOA and random guessing. Consider a depth-$p$ trained QAOA ansatz sampling $N$ possible solutions labeled by an integer $x \in [N]$ and ordered by decreasing order of probability. Denoting by $\left(q_x\right)_{x \in [N]}$ the probability distributions of the solutions ($q_x$ is decreasing by definition), let
\begin{align}
	P_{\mathrm{random}}(q) & := 1 - \left(1 - \frac{1 + \min\left\{x \in [N]\,:\,\sum\limits_{0 \leq y \leq x}q_y \geq q\right\}}{N}\right)^p.
\end{align}
The above is easily seen to be the probability of obtaining a solution among the level $q$ quantile after $p$ attempts of random guessing. On the other hand, the probability of obtaining such a solution by sampling from the QAOA ansatz is, by definition, $q$. The motivation behind this comparison is that $p$ queries (either classical or quantum) are made to the cost function in both cases. One may then consider the ratio of success probabilities $\frac{q}{P_{\mathrm{random}}(q)}$; it is greater than 1 iff. QAOA gives an advantage over random guessing. This quantity is represented in figure \ref{fig:ala_peptide_qaoa_vs_random_guessing} for $p \in \{2, 3, 8, 62\}$.
The figures show that QAOA outperforms random guessing only for small quantiles $q$, with a very mild maximum ratio $6.6$. Furthermore, note that our metric does not account for the queries to the cost function required to train the ansatz. Factoring these queries in would effectively void the advantage; on the other hand, the training might be avoided provided one could guess good enough (not necessarily optimal) variational parameters. Finally, all these results ultimately depend on the parameter optimization protocol described in section \ref{sec:ala_peptide_variational_optimization}; it may be that the variational parameters we found are highly suboptimal, explaining the modest performance of QAOA on the problem instances considered.

\section{Conclusion}

In this work, we investigated the feasibility of sampling low-energy lattice-based peptides using a well-studied variational quantum algorithm: the Quantum Approximate Optimization Algorithm (QAOA). The choice of this cost-function-dependent algorithm was motivated by its better trainability compared to cost-function-agnostic algorithms considered in earlier works. The performance of QAOA was first evaluated on a highly simplified version of the peptide folding problem, reduced to sampling a self-avoiding walk. The algorithm showed promising results in this setting, though uncertainties remain as to the fairness of the sampling and the scaling of the success probability for large (not classically simulatable) problem sizes. In contrast, there is strong empirical evidence that QAOA can be efficiently trained on this simple formulation of the problem, addressing an important practical challenge of the algorithm. QAOA achieved more mixed results on the full lattice-based peptide folding problem. While it still produces valid (self-avoiding) conformations with high probability, it struggles to find low energy instances among these even at high depth ($\sim 100$) and for a very small peptide (4 amino acids). These results may either point to instrinsic limitations of QAOA (reachability deficit at low depth) or to a shortcoming of our training protocol. These negative results could indicate that QAOA should be applied to constraint satisfaction problems rather than discretized continuous or mixed optimization problems.

The formulation of lattice-based peptide folding used in this work is limited and could be generalized in several ways. For instance, it is not strictly necessary for the atoms to lie on a tetrahedral lattice and one may consider different bond lengths and angles for consecutive pairs of atoms. One could also increase the number of degrees of freedom per bond (e.g.: more dihedral angles or variable bond lengths); such generalizations may be hard to investigate on current quantum emulators (due to higher qubit requirements) but should be considered when large-scale fault-tolerant quantum computers are available. Finally, in a different direction to the quantitative energetic view adopted in this work, it may be worth applying quantum optimization algorithms to \textit{qualitative} peptide scoring functions derived from knowledge-based approaches.

\subsection*{Acknowledgements}

This project has received funding from the European Research Council (ERC) under the European Union's Horizon 2020 research and innovation programme (grant agreement No.\ 817581) and was supported by the EPSRC Centre for Doctoral Training in Delivering Quantum Technologies, grant ref. EP/S021582/1. Google Cloud credits were provided by Google via the EPSRC Prosperity Partnership in Quantum Software for Modeling and Simulation (EP/S005021/1).

We thank Martin Strahm and Mari\"{e}lle van de Pol for overseeing the Roche Quantum Computing Taskforce and this project.


\printbibliography

\appendix

\section{Extra figures}
\label{sec:extra_figures}

\subsection{Sampling low-energy peptide conformations with QAOA}
\label{appendix:figures_small_protein}

\paragraph{Conversion between relative and absolute turn encoding.} In table \ref{tab:relative_to_absolute_encoding}, we explicitly specify the conversion between the relative and absolute turn-based encodings introduced in section \ref{sec:encoding_details}.
\begin{table}[!htbp]
    \centering
    \begin{tabular}{c|c|c}
        Previous turn (absolute) & Current turn (relative) & Current turn (absolute)\\
        \hline
        0 & 0 & 1\\
        \hline
        0 & 1 & 2\\
        \hline
        0 & 2 & 3
    \end{tabular}

    \vspace*{10px}

    \begin{tabular}{c|c|c}
        Previous turn (absolute) & Current turn (relative) & Current turn (absolute)\\
        \hline
        1 & 0 & 2\\
        \hline
        1 & 1 & 3\\
        \hline
        1 & 2 & 0
    \end{tabular}

    \vspace*{10px}

    \begin{tabular}{c|c|c}
        Previous turn (absolute) & Current turn (relative) & Current turn (absolute)\\
        \hline
        2 & 0 & 3\\
        \hline
        2 & 1 & 0\\
        \hline
        2 & 2 & 1
    \end{tabular}

    \vspace*{10px}

    \begin{tabular}{c|c|c}
        Previous turn (absolute) & Current turn (relative) & Current turn (absolute)\\
        \hline
        3 & 0 & 0\\
        \hline
        3 & 1 & 1\\
        \hline
        3 & 2 & 2
    \end{tabular}
    \caption{Relative and absolute turn-based encodings}
    \label{tab:relative_to_absolute_encoding}
\end{table}

\begin{figure}[!tbp]
	\centering
	\includegraphics[width=0.5\textwidth]{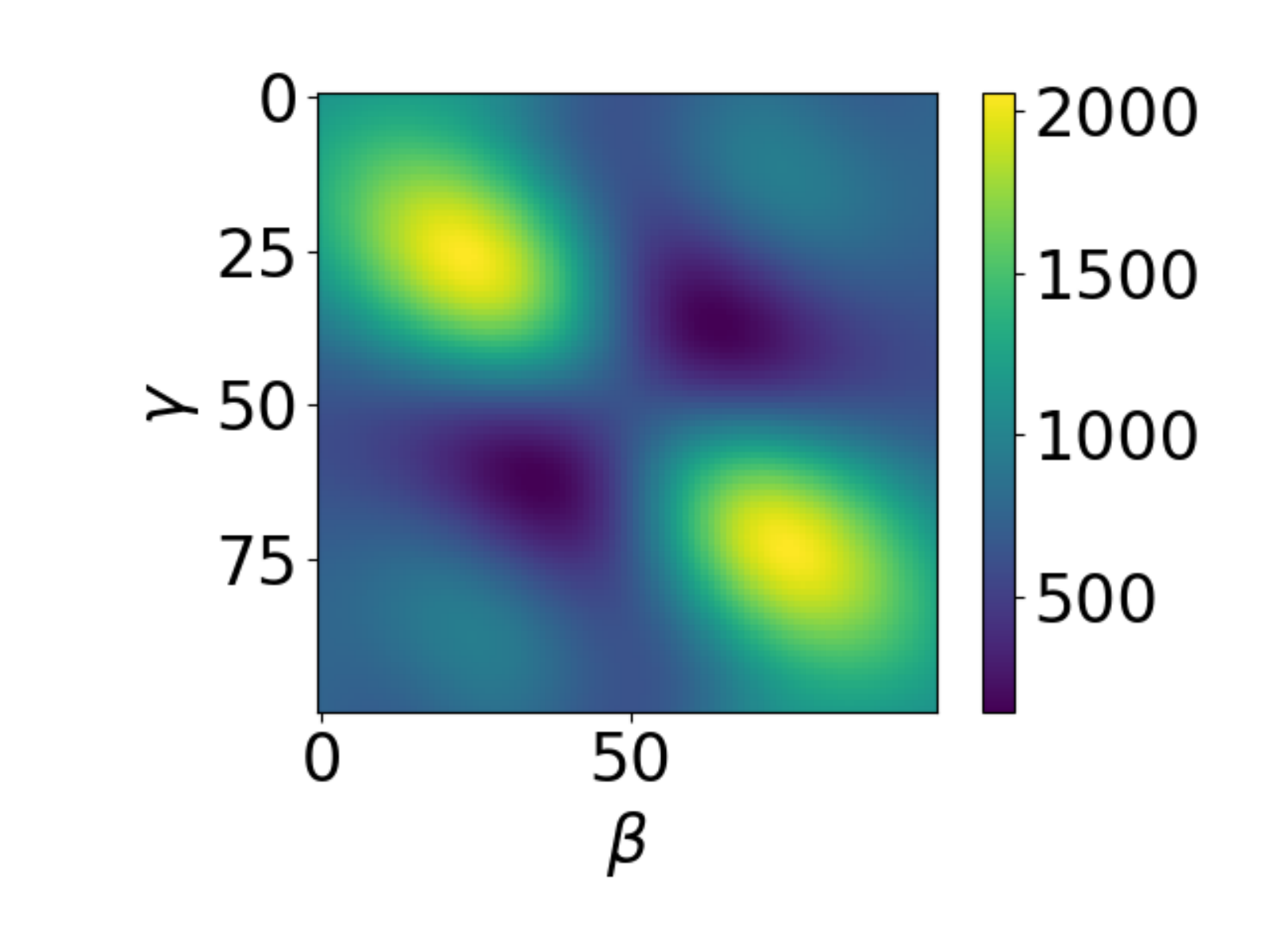}
	\caption{$p = 1$ QAOA landscape for the alanine dipeptide. $\beta$ and $\gamma$ are the hyperparameters of the quantum circuit ansatz. The represented cost function is the energy in equation \ref{eq:ala_peptide_full_potential} (expressed in $\textrm{kcal.mol}^{-1}$) averaged over conformations sampled from the ansatz. The clash penalization parameter $\lambda$ is set to $1000\,\textrm{kcal.mol}^{-1}$.}
	\label{fig:qaoa_landscape_p=1}
\end{figure}

\paragraph{$p = 1$ QAOA landscape}
We provide on figure \ref{fig:qaoa_landscape_p=1} a view of the $p = 1$ QAOA landscape after the parameter restriction and rescaling described in section \ref{sec:ala_peptide_variational_optimization} have been applied.

\begin{figure}[!tbp]
	\begin{subfigure}{0.32\textwidth}
		\centering
		\includegraphics[width=\textwidth]{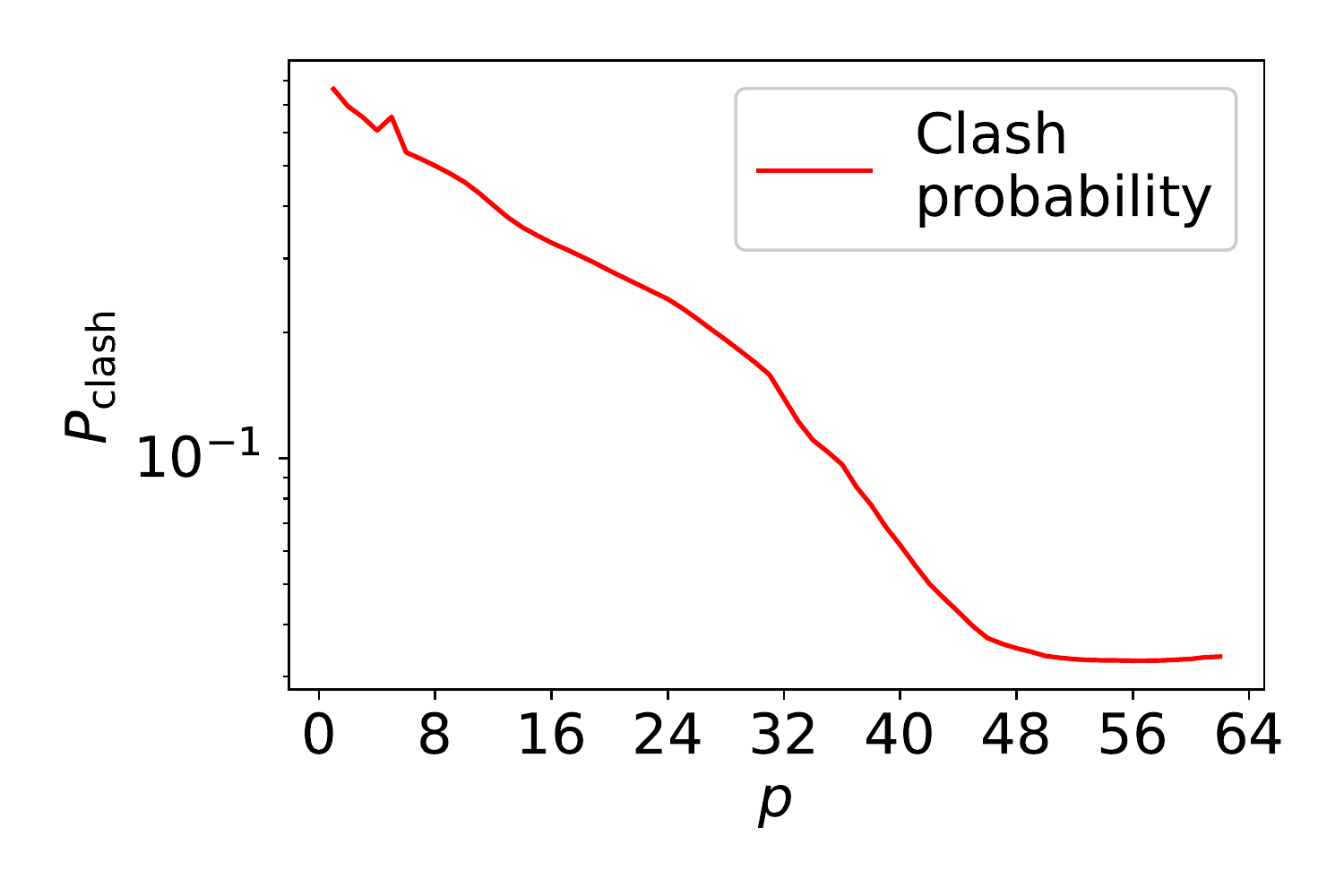}
		\caption{Random initialization}
		\label{subfig:ala_peptide_clash_probability_random}
	\end{subfigure}
	\begin{subfigure}{0.35\textwidth}
		\centering
		\includegraphics[width=\textwidth]{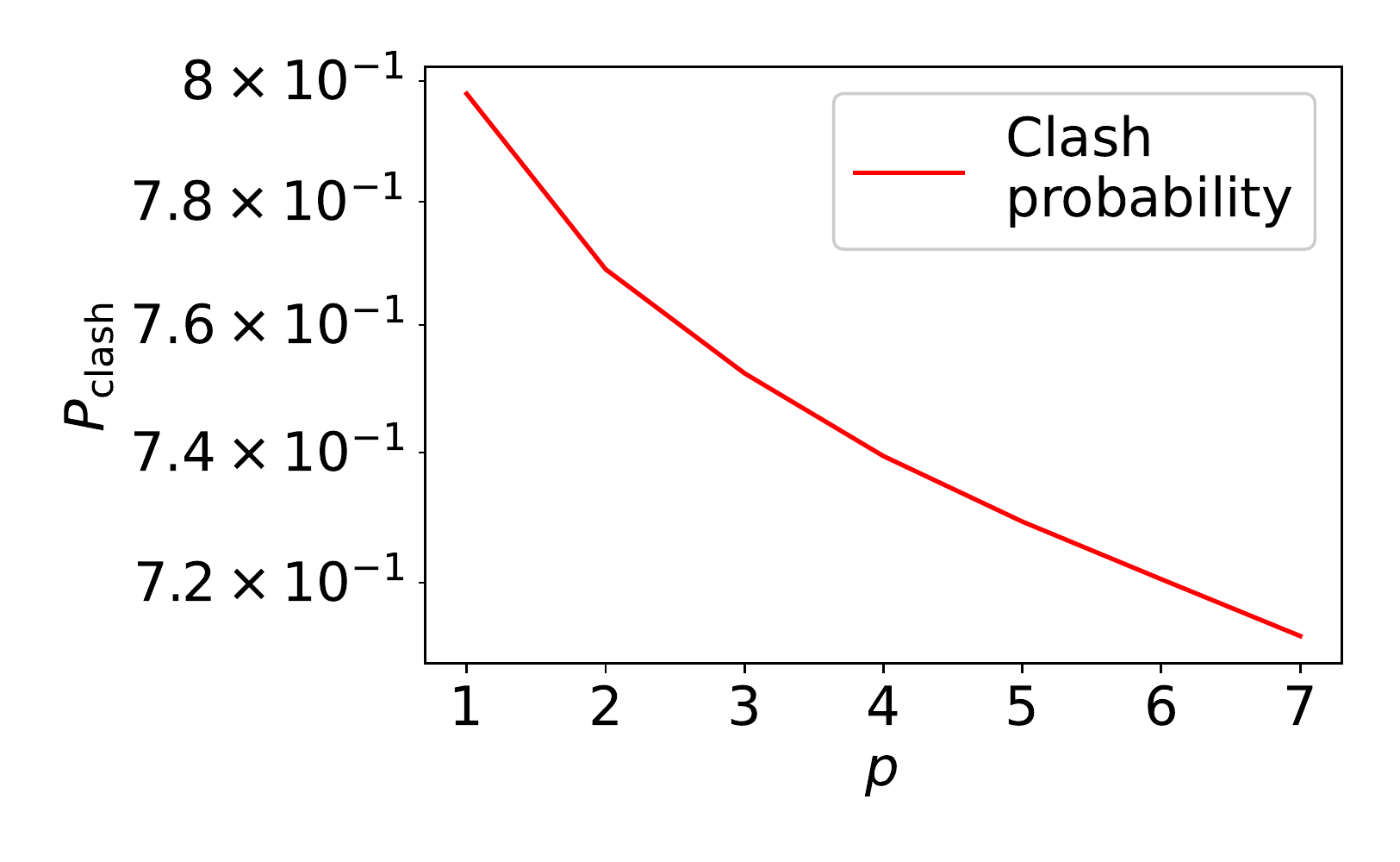}
		\caption{Quantum annealing schedule}
		\label{subfig:ala_peptide_clash_probability_adiabatic}
	\end{subfigure}
	\begin{subfigure}{0.32\textwidth}
		\centering
		\includegraphics[width=\textwidth]{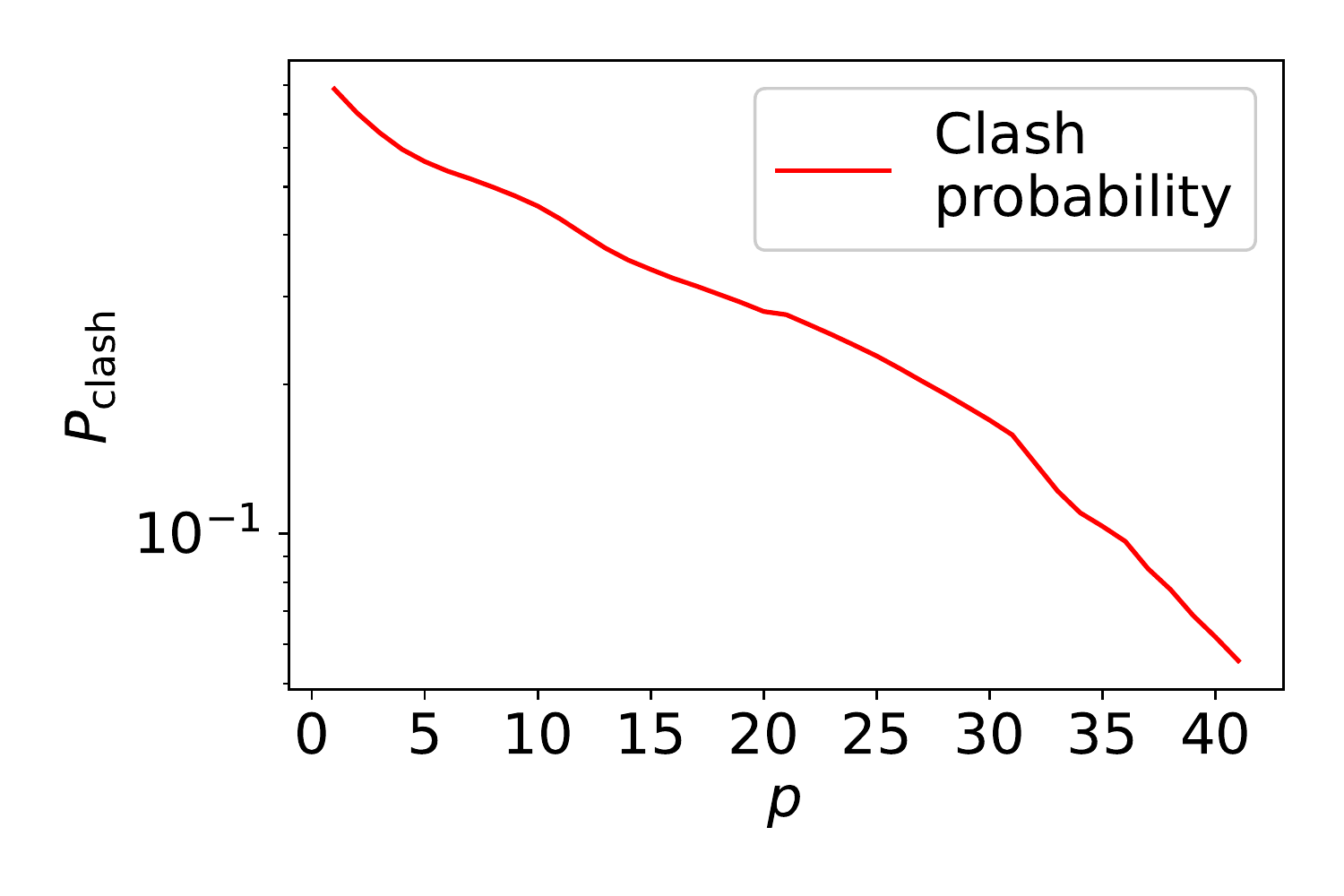}
		\caption{Optimized from quantum annealing schedule}
		\label{subfig:ala_peptide_clash_probability_init_adiabatic}
	\end{subfigure}
	\caption{Clash probability vs.\ ansatz depth for alanine tetrapeptide problem from section \ref{sec:sampling_low_energy_conformations}}
	\label{fig:ala_peptide_clash_probability}
\end{figure}

\paragraph{Suppression of clashes.} We provide here additional numerical evidence, besides figure \ref{fig:alanine_peptide_energy_optimization} from section \ref{sec:ala_peptide_results}, showing QAOA is particularly efficient at suppressing clashes and that this feature explains most of its performance. This is illustrated in figure \ref{fig:ala_peptide_clash_probability}, where the probability of sampling a conformation with clashes is represented against the ansatz depth. The figures are compatible with an suppression of the clash probability scaling exponentially with the ansatz depth. Therefore, the improvement of the energy with increasing $p$ observed in section  \ref{sec:ala_peptide_results}, figure \ref{fig:alanine_peptide_energy_optimization} is more attributable to the decline of clashes than to the decrease of the Lennard-Jones energy conditioned on the absence of clash. In other words, it appears QAOA is solving a relatively uninteresting constraint satisfaction problem (generating a physically valid protein conformation) rather than the actual protein folding problem.

\begin{figure}[!tbp]
	\centering
	\includegraphics[width=0.5\textwidth]{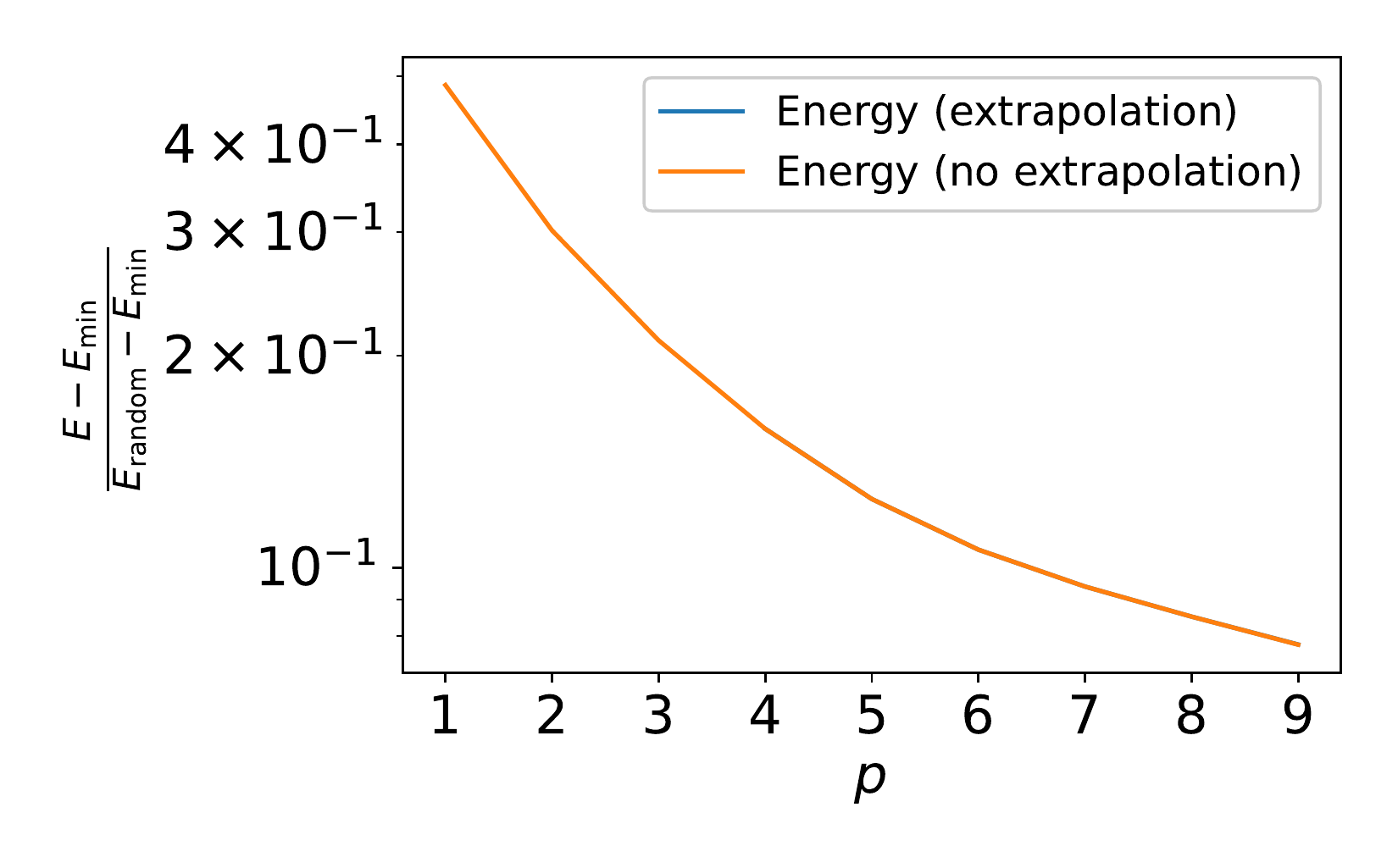}
	\caption{Energy achieved by QAOA on alanine tetrapeptide problem from section \ref{sec:sampling_low_energy_conformations}, with or without extrapolated initialization. The curve without extrapolation (orange) was obtained by selecting the best of 50 optimization attempts, each of which started with uniformly random parameters, for each $p$. The curve with extrapolation (blue) is common with the curve without extrapolation for $1 \leq p \leq 3$. However, for each $p \geq 4$, parameters were optimized (with a single attempt since the optimizer is deterministic) starting from a linear extrapolation of the optimal parameters found at level $p - 1$.}
	\label{fig:ala_peptide_extrap_vs_no_extrap}
\end{figure}

\paragraph{Variational parameters extrapolation.} We illustrate here the merit of variational parameters extrapolation, similar to the self-avoiding walk example (figure \ref{fig:saw_extrapolation} section \ref{sec:saw_results}). Figure \ref{fig:ala_peptide_extrap_vs_no_extrap} compares the QAOA energy achieved by for $4 \leq p \leq 9$ starting either from uniformly random or from extrapolated parameters. The figure shows that initialization by extrapolation performs no worse than 50 uniformly random initialization attempts.

\subsection{Sampling self-avoiding walks with QAOA}
\label{appendix:figures_saw_qaoa}

\begin{figure}[!tbp]
	\centering
	\begin{subfigure}{0.32\textwidth}
		\includegraphics[width=\textwidth]{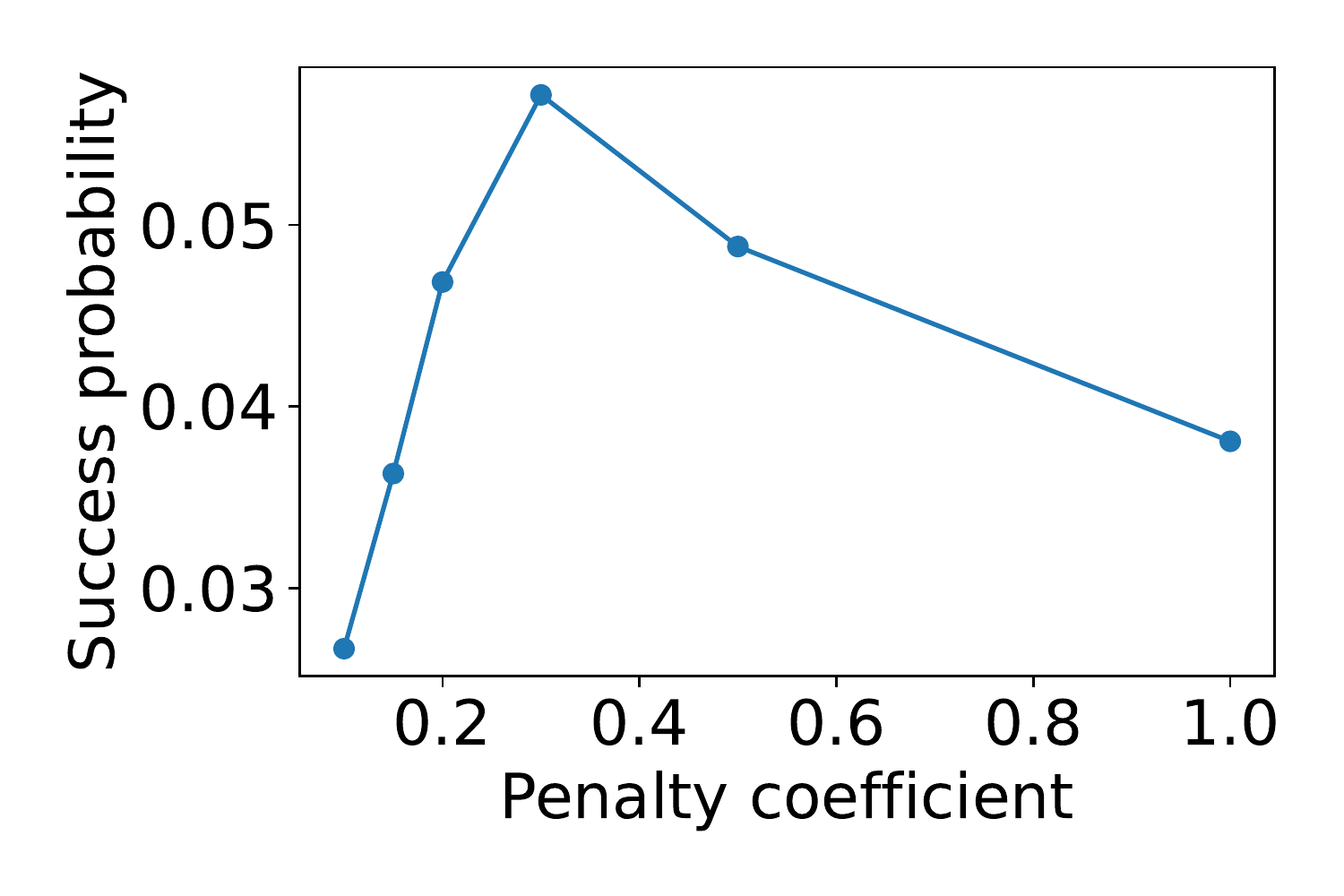}
		\caption{Level $p = 1$ QAOA}
	\end{subfigure}
	\begin{subfigure}{0.32\textwidth}
		\includegraphics[width=\textwidth]{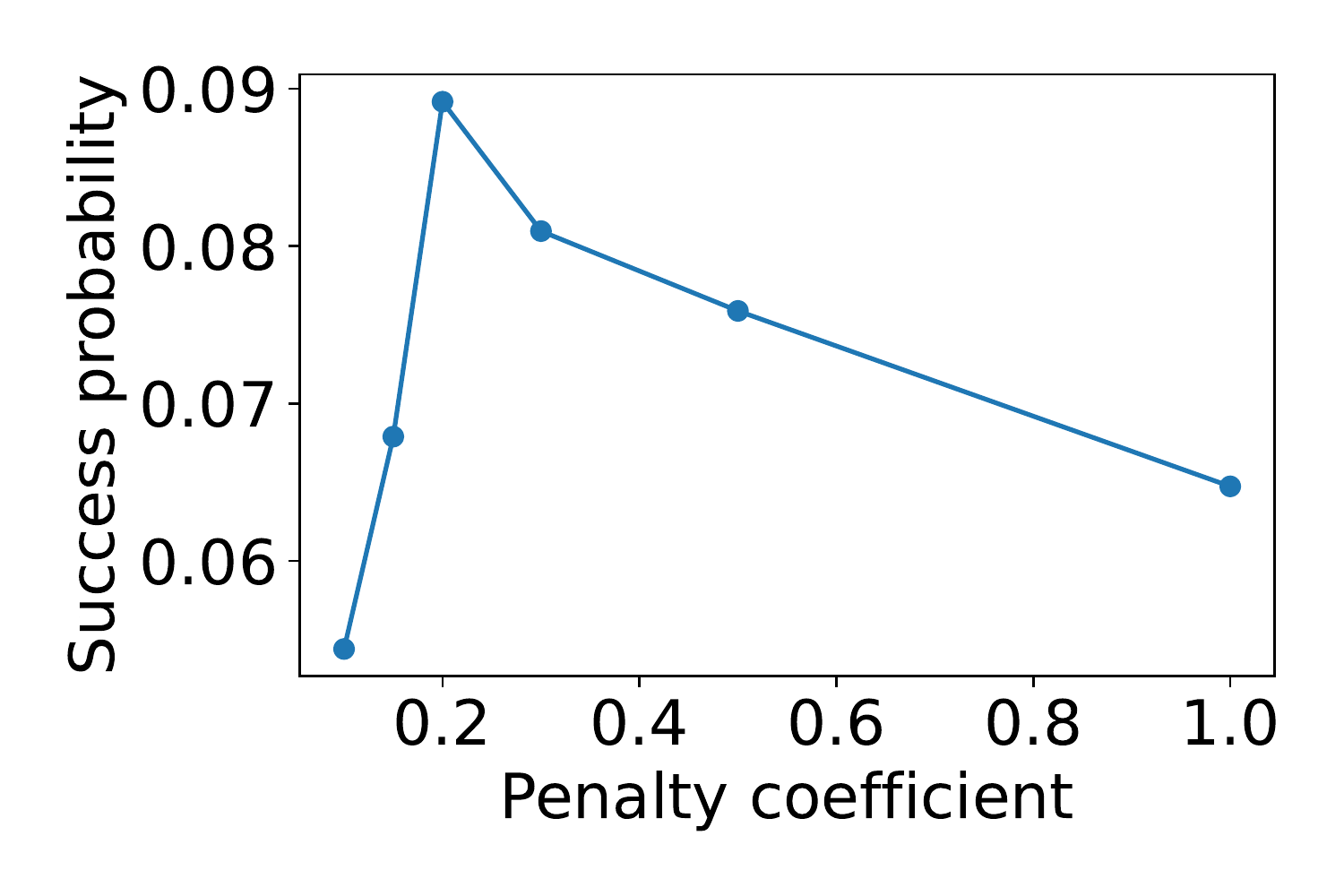}
		\caption{Level $p = 2$ QAOA}
	\end{subfigure}
	\begin{subfigure}{0.32\textwidth}
		\includegraphics[width=\textwidth]{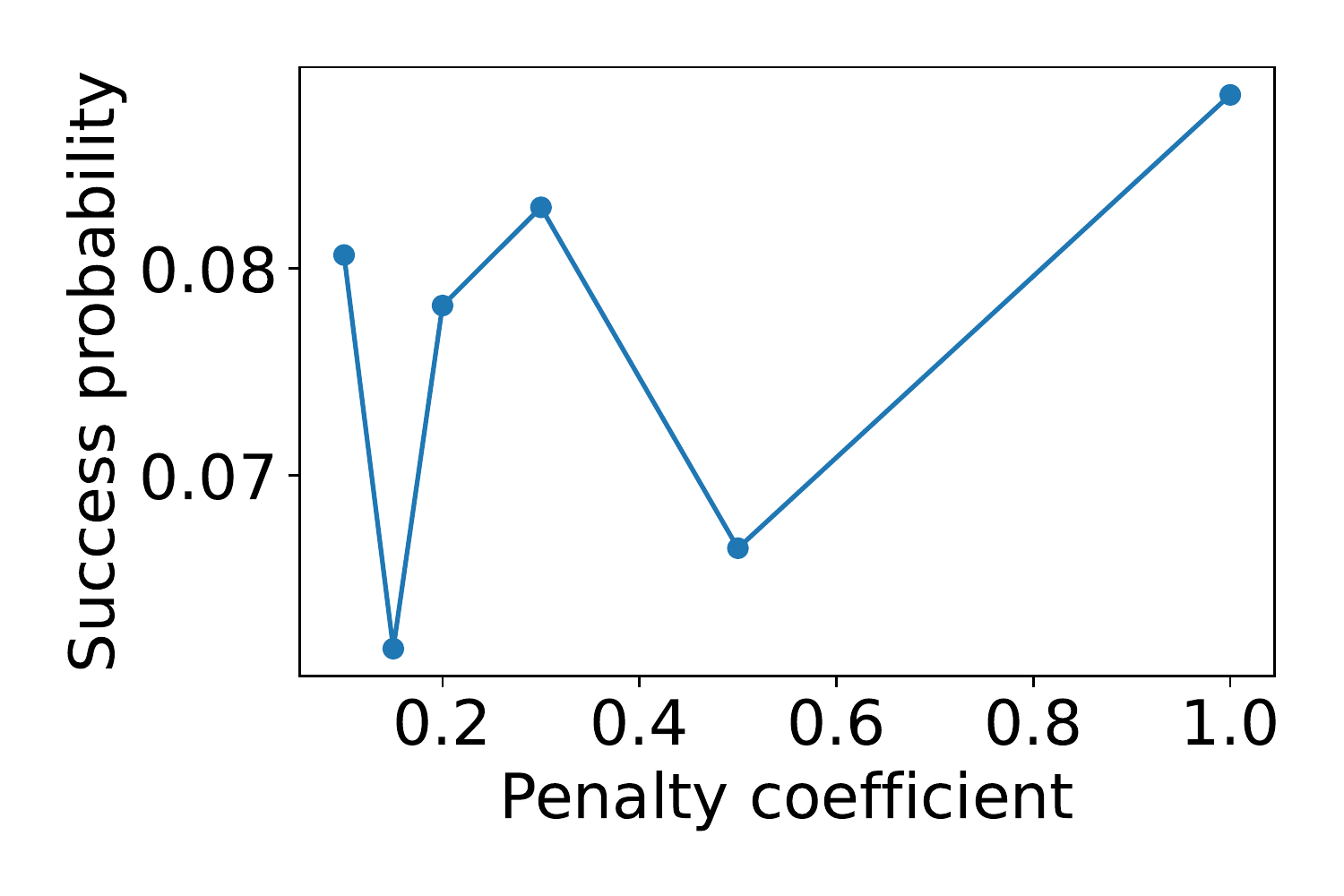}
		\caption{Level $p = 3$ QAOA}
	\end{subfigure}
	\caption{Penalty tuning for a 6-step self-avoiding walk (Grover mixer, absolute turn encoding)}
	\label{fig:penalty_tuning_6_steps}
\end{figure}

\begin{figure}[!tbp]
	\centering
	\begin{subfigure}{0.32\textwidth}
		\includegraphics[width=\textwidth]{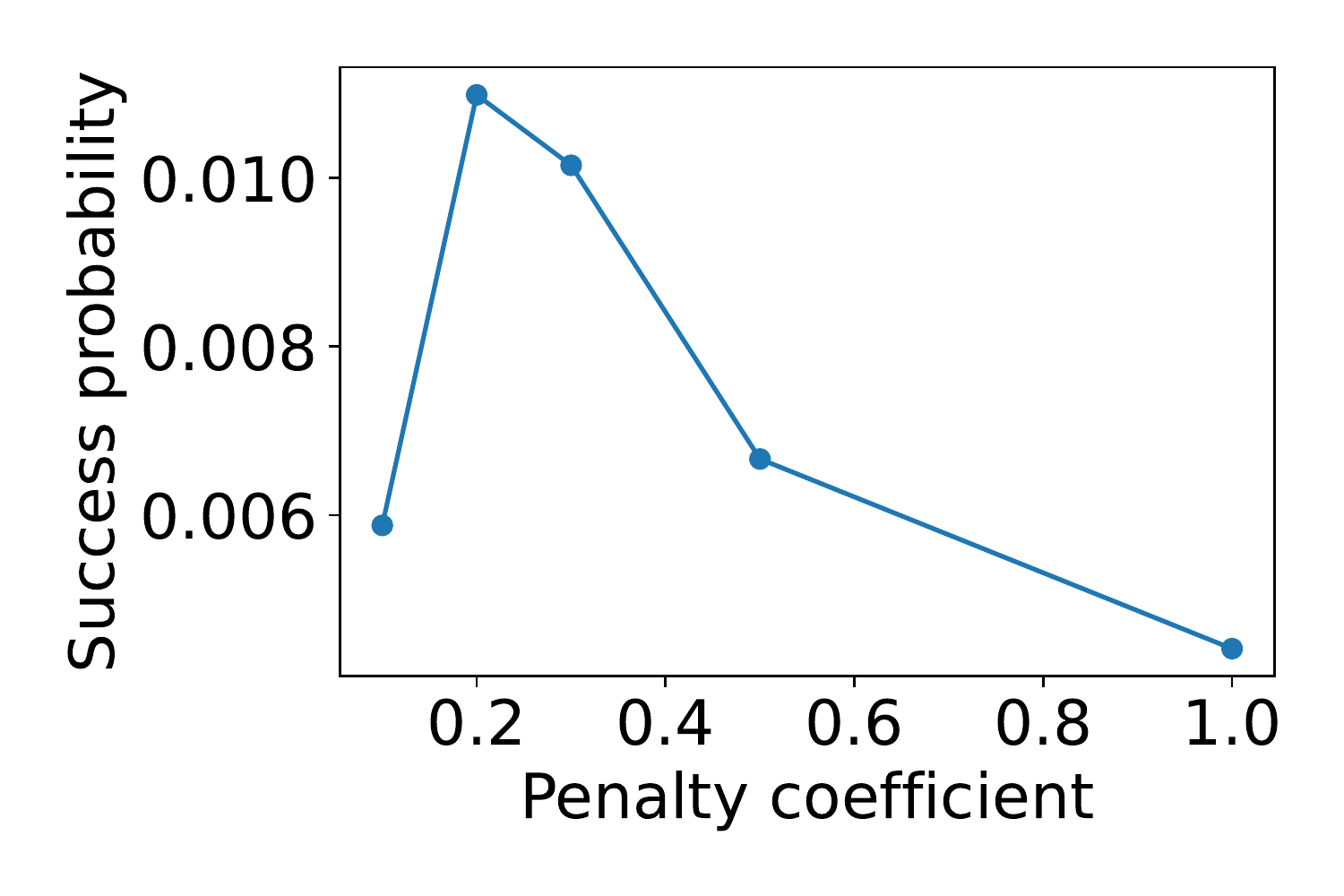}
		\caption{Level $p = 1$ QAOA}
	\end{subfigure}
	\begin{subfigure}{0.32\textwidth}
		\includegraphics[width=\textwidth]{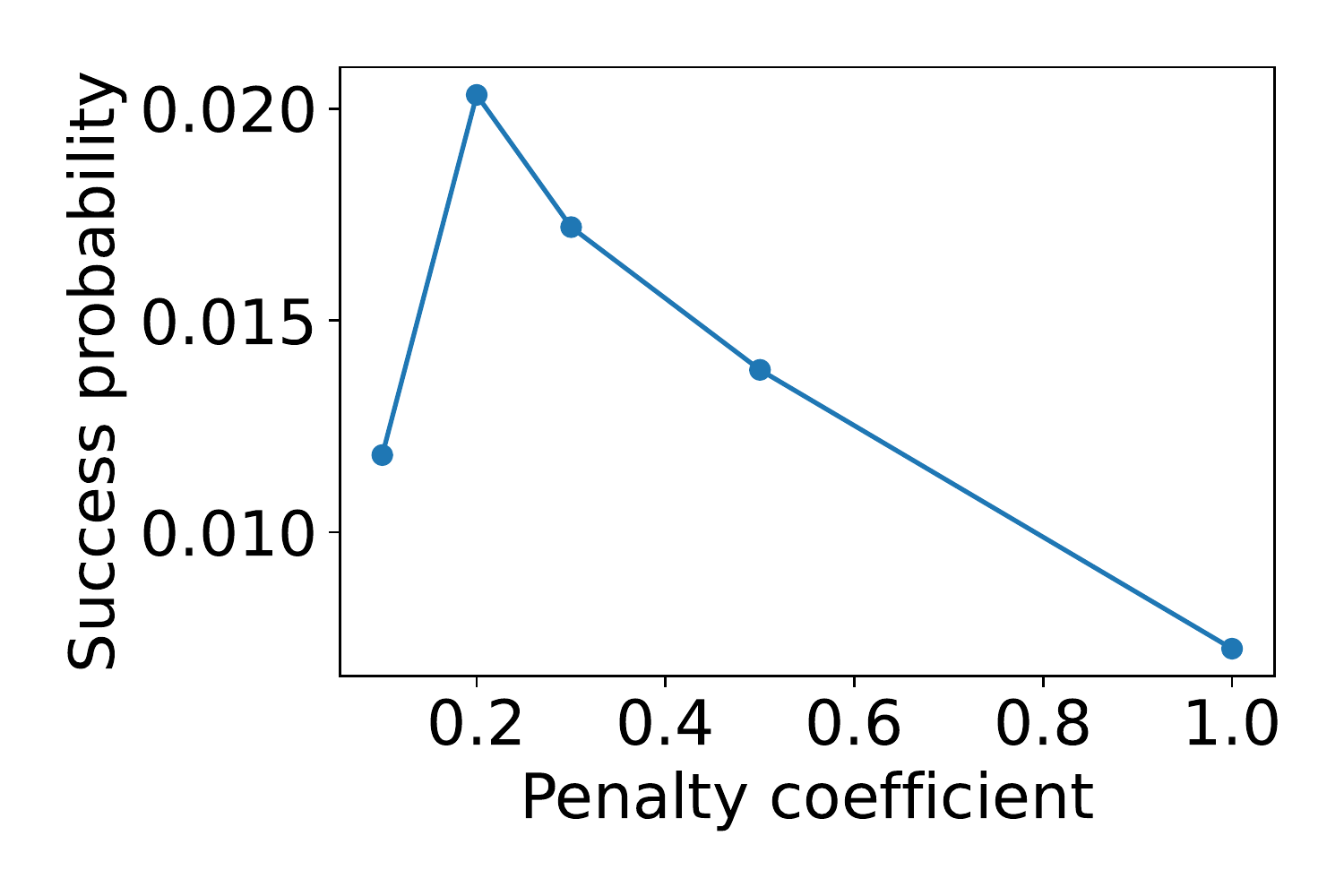}
		\caption{Level $p = 2$ QAOA}
	\end{subfigure}
	\begin{subfigure}{0.32\textwidth}
		\includegraphics[width=\textwidth]{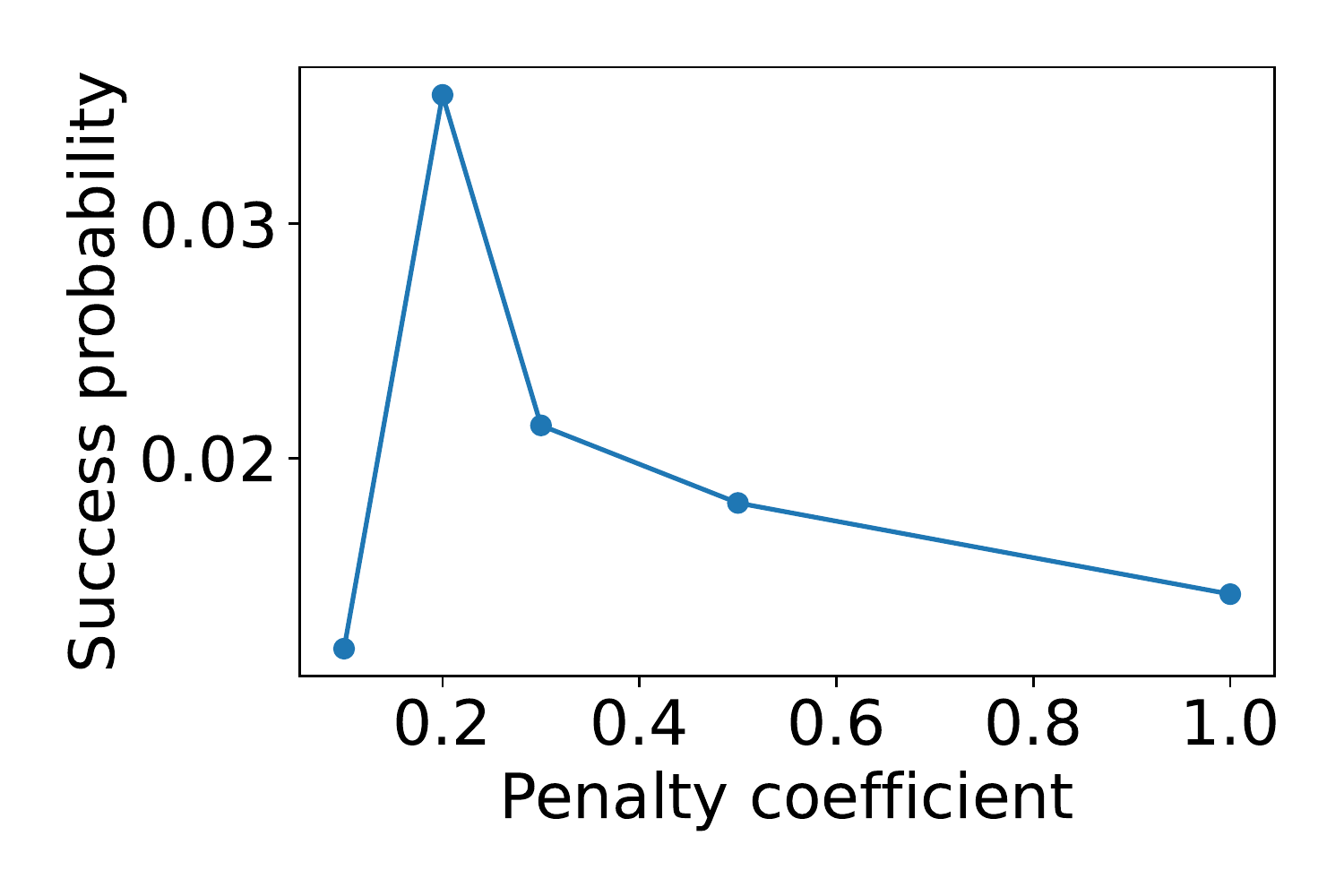}
		\caption{Level $p = 3$ QAOA}
	\end{subfigure}
	\caption{Penalty tuning for a 10-step self-avoiding walk (Grover mixer, absolute turn encoding)}
	\label{fig:penalty_tuning_10_steps}
\end{figure}

\paragraph{Penalty tuning.}
Figure \ref{fig:penalty_tuning_6_steps} (resp. \ref{fig:penalty_tuning_10_steps}) illustrates the penalty tuning protocol described in section \ref{sec:saw_experiments} for QAOA applied to a 6-step (resp. 10-step) walk. It represents the probability of sampling a self-avoiding loop for different values of the penalty hyperparameter.

\begin{figure}[!tbp]
	\centering
	\begin{subfigure}{0.47\textwidth}
		\centering
		\includegraphics[width=\textwidth]{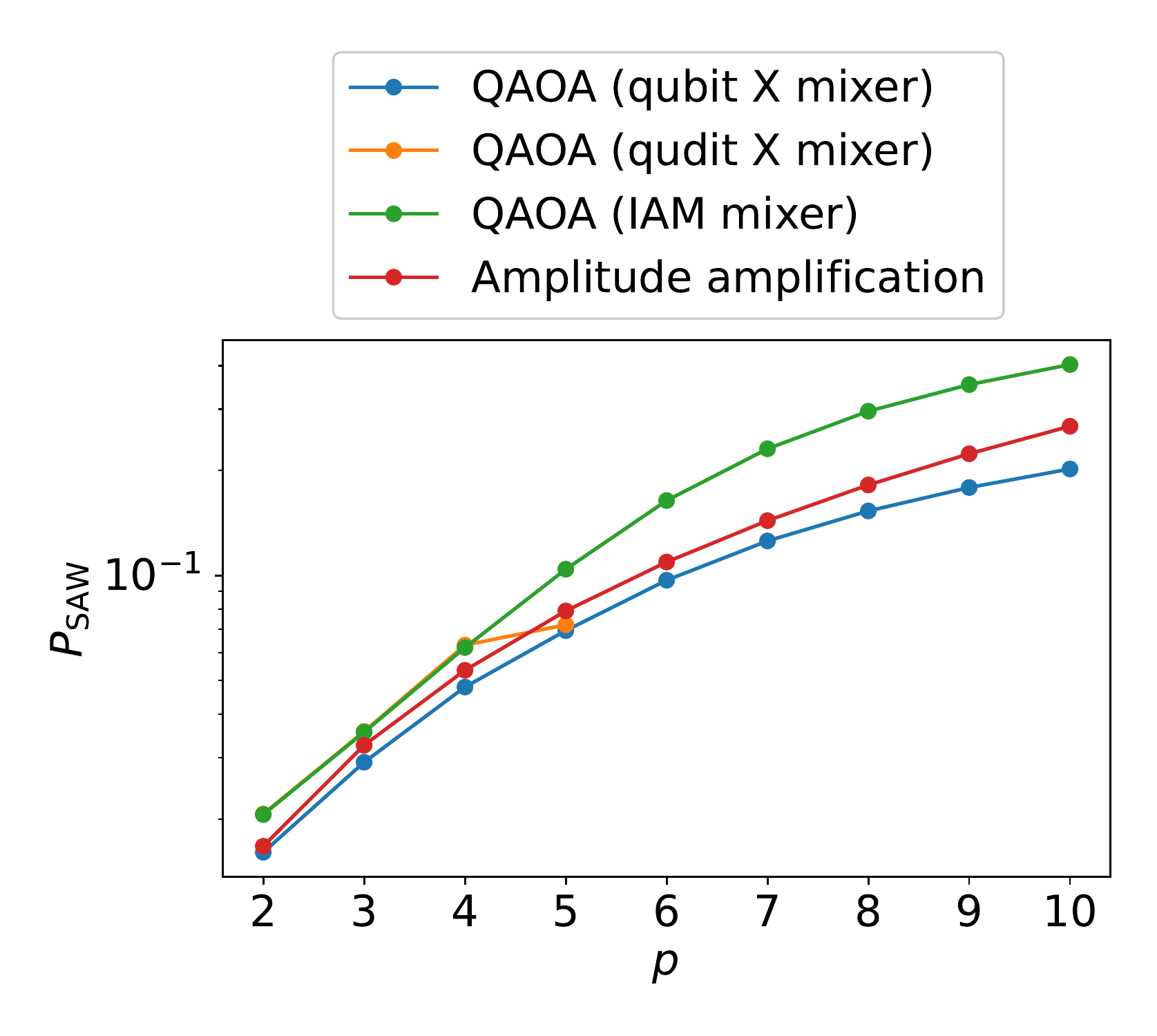}
		\caption{Backtracking encoding}
		\label{subfig:mixer_comparison_backtracking}
	\end{subfigure}
	\begin{subfigure}{0.47\textwidth}
		\centering
		\includegraphics[width=\textwidth]{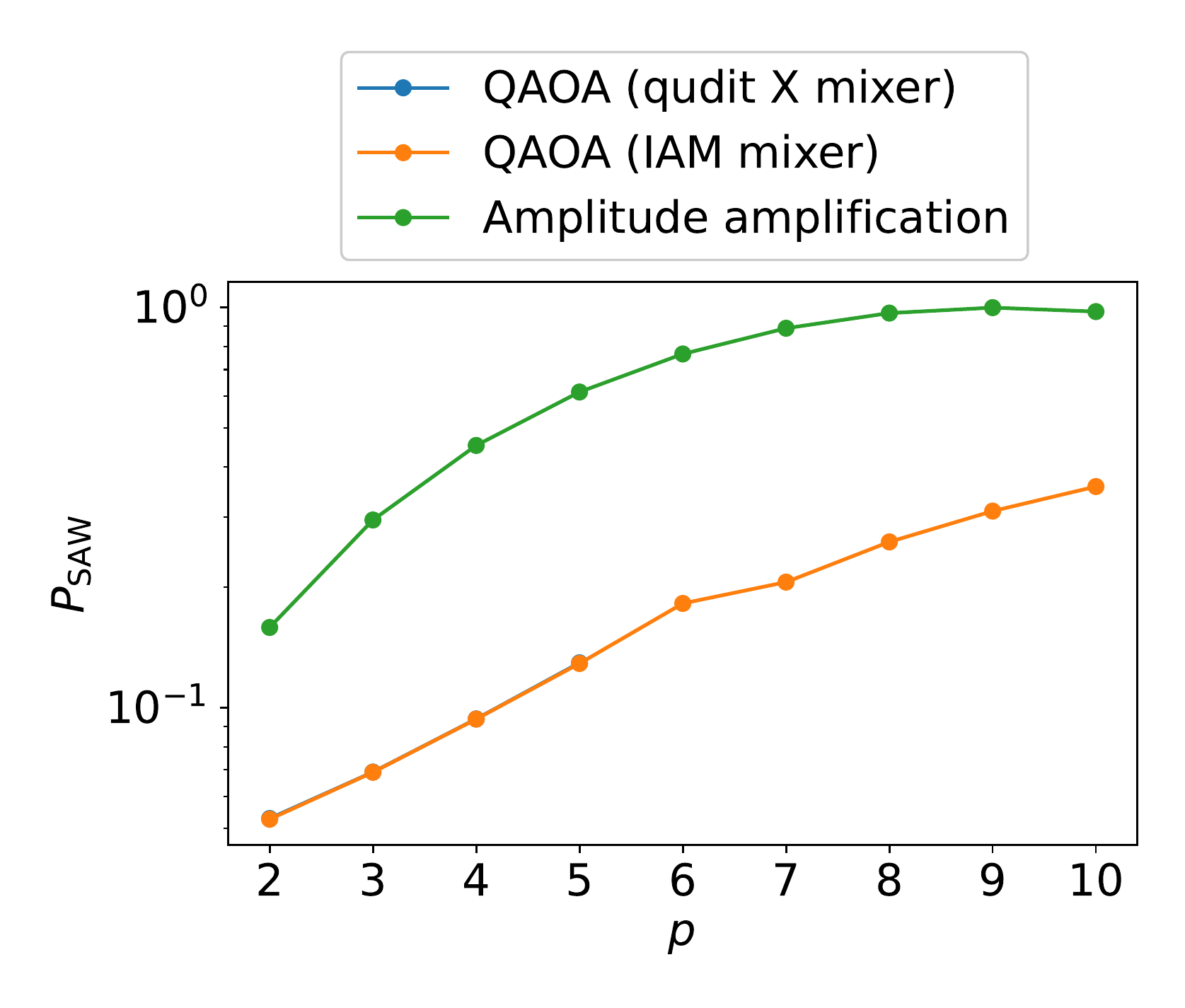}
		\caption{Non-Backtracking encoding}
		\label{subfig:mixer_comparison_non_backtracking}
	\end{subfigure}
	\caption{Probability of sampling a self-avoiding loop (10 steps) vs. ansatz depth for different choices of mixers}
	\label{fig:mixer_comparison}
\end{figure}

\paragraph{Mixer and encoding comparison.}
Figure \ref{fig:mixer_comparison}, which complements tables \ref{tab:mixer_comparison_backtracking} and \ref{tab:mixer_comparison_non_backtracking} section \ref{sec:saw_results}, compares the success probability in sampling a 10-step self-avoiding walk starting from a uniform superposition of non self-avoiding walks and applying either amplitude amplification or QAOA. For QAOA, the backtracking and non-backtracking encodings (see section \ref{sec:qaoa_saw_encoding_hamiltonian}) and different mixers (see section \ref{sec:saw_mixers}) are compared.

\begin{figure}[!tbp]
	\centering
	\includegraphics[width=0.65\textwidth]{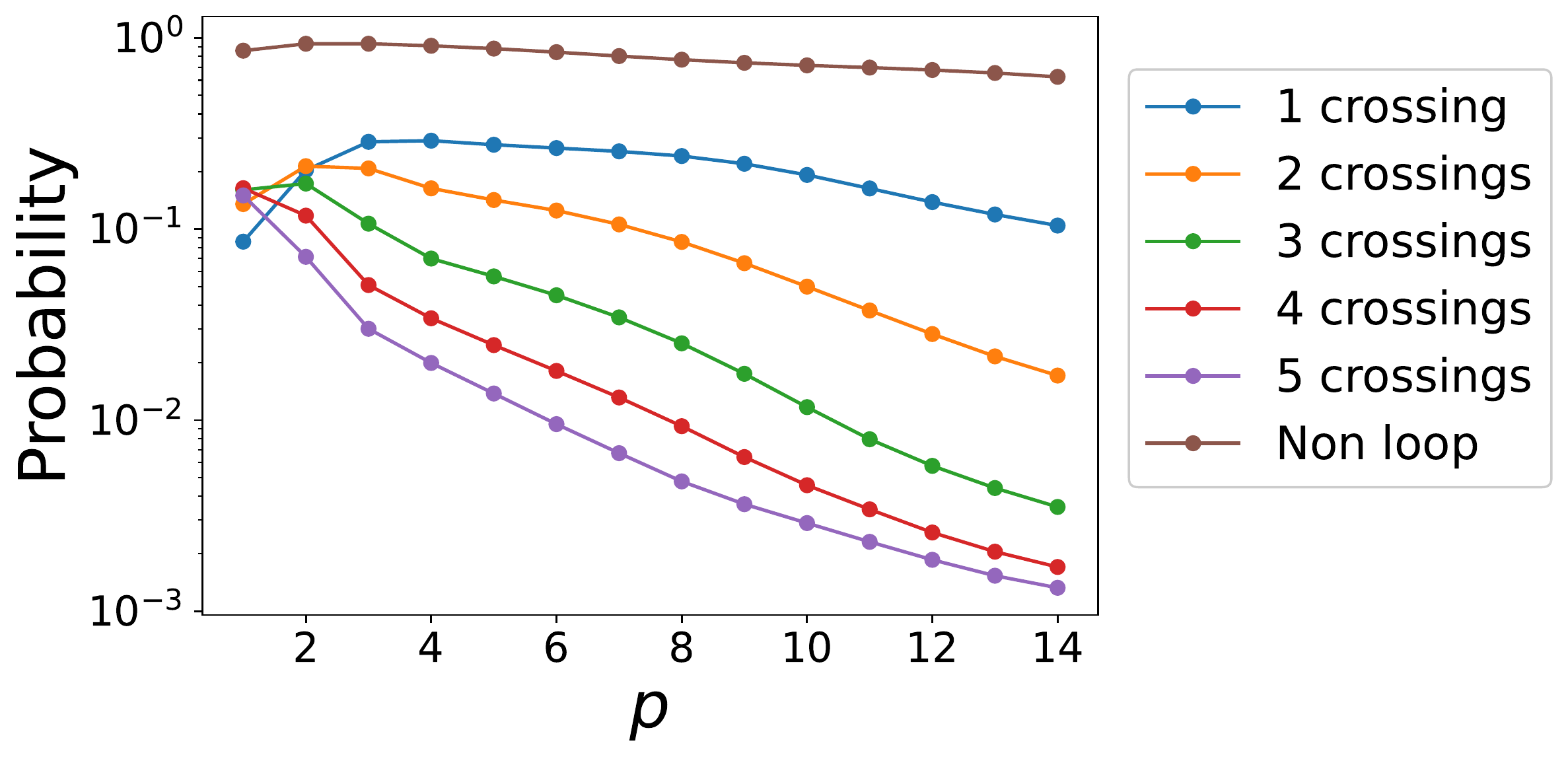}
	\caption{Probability distribution of number of self-crossings and probability of non-loop for 16-step walks sampled from level-$p$ QAOA (inversion about the mean encoding). Higher numbers ($ > 5$) of crossings are possible but not represented here for space reasons; the curves lie below those represented.}
	\label{fig:proba_dist_clashes_vs_p}
\end{figure}

\paragraph{Analysis of success probability for 16-step walk.} The success probability is further analyzed for the 16-step walk in figure \ref{fig:proba_dist_clashes_vs_p}, giving the probability of the walk having $k$ self-crossings for $1 \leq k \leq 5$ as $p$ increases. The figure shows that at low $p$, high number of crossings are exponentially suppressed but low number of crossings also increase. Past $p = 3$, all types of crossings decline but the large $p$ scaling is unclear.

\end{document}